\documentclass[12pt]{article}
\usepackage{graphicx}
\graphicspath{{img/}}
\usepackage{subcaption}
\usepackage{amsmath}
\usepackage{amssymb}
\usepackage{csquotes} % \textquote{ } -- for quotes
\usepackage{physics}
\usepackage{slashed}

\usepackage{mathtools}
\usepackage{amsfonts}
\numberwithin{equation}{section}

\usepackage{xcolor}
\usepackage{hyperref}
\usepackage{placeins} % \FloatBarrier

\newcommand{\beq}{\begin{equation}}   
\newcommand{\eeq}{\end{equation}}

\newcommand{\pt}{\partial}

\newcommand{\Zc}{{\mathcal Z}}
\newcommand{\Wc}{{\mathcal W}}

\def\none{${\mathcal N}=1\;$}

\begin{document}

\hypersetup{%
	%colorlinks=false,% hyperlinks will be black
	linkbordercolor=blue,% hyperlink borders will be =color
	%pdfborderstyle={0 0 0.1}% 
}

\begin{titlepage}

\begin{flushright}
FTPI-MINN-25-11 \\ UMN-TH-4506/25
\end{flushright}

%\vspace{2mm}

% Evgenii Ievlev and Mikhail Shifman
% E. Ievlev and M. Shifman
% Degenerate kinks and kink-instantons in two-dimensional scalar field models with $\mathcal{N}=1$ and $\mathcal{N}=2$ supersymmetry
% Degenerate kinks and kink-instantons in two-dimensional scalar field models with N=1 and N=2 supersymmetry

\begin{center}
	\Large{{\bf \boldmath
		Degenerate kinks and kink-instantons \\ in two-dimensional scalar field models \\ with $\mathcal{N}=1$ and $\mathcal{N}=2$ supersymmetry
		}}

\vspace{5mm}
	
{\large  \bf Evgenii Ievlev and Mikhail Shifman}

\end{center}
\begin{center}
{\it  Department of Physics,
University of Minnesota,
Minneapolis, MN 55455}\\[5pt]
{\it  William I. Fine Theoretical Physics Institute,
University of Minnesota,
Minneapolis, MN 55455}\\
\end{center}

\vspace{5mm}

\begin{center}
{\large\bf Abstract}
\end{center}

Models with classically degenerate vacua often support quasiclassical configurations of nontrivial topology. In (0+1)-dimensional quantum mechanics with a double-well potential, for example, instantons induce mixing between the two perturbative ground states in the purely bosonic case, while in the supersymmetric version, the tunneling amplitude is suppressed.

In this work, we investigate (1+1)-dimensional models featuring classically Bogomol'nyi-Prasad-Sommerfield saturated kinks with degenerate masses and identical topology. Recent studies suggest that such kinks may undergo mixing mediated by scalar-field instantons. We analyze this phenomenon in a supersymmetric framework and demonstrate that, whereas mixing indeed occurs in the bosonic theory, the presence of fermionic zero modes in the supersymmetric case leads to the vanishing of the transition amplitude. To illustrate these results, we examine two examples featuring Wess-Zumino models with two and four supercharges. The latter example is motivated by the Affleck-Dine-Seiberg superpotential. We also present a number of developments of instanton calculus in the case of instantons  in kink backgrounds.

\end{titlepage}

{
%\footnotesize
%\small
\setcounter{tocdepth}{2}
\tableofcontents
}

%\newpage

\section{Introduction}
\label{intro}

Quasiclassical nonperturbative effects have proved to be deeply woven into the fabric of non-Abelian gauge theories. Suffice it to mention
't Hooft-Polyakov monopoles 
\cite{tHooft:1974kcl}, instantons \cite{Belavin:1975fg}, sphalerons \cite{Dashen:1974ci,Klinkhamer:1984di}, etc., which resulted in such phenomena as proton decay catalysis \cite{Rubakov:1982fp,Callan:1982ac} and axions as a possible solution to the problem of CP conservation in QCD and possible candidates for dark matter \cite{Weinberg:1977ma,Wilczek:1977pj,Shifman:1979if,Kim:1979if,Sikivie:1983ip}. The advent of supersymmetry (SUSY), with its powerful methods, allowed one to advance from the quasiclassical approximation to exact results at strong coupling. The culmination of this story -- but not the end of it --  was the first analytic proof of confinement in
${\mathcal N}=2$ super-Yang-Mills by Seiberg and Witten \cite{Seiberg:1994rs}. Many new exact results were obtained for the Bogomol'nyi-Prasad-Sommerfield (BPS) protected objects, including extended objects, in the past two decades. As some unexpected examples, let us mention domain walls in ${\mathcal N}=1$ super-Yang-Mills \cite{Dvali:1996xe} or non-Abelian vortex tubes \cite{Auzzi:2003fs,Shifman:2004dr,Hanany:2004ea}.

For obvious reasons, investigations in this direction were focused on non-Abelian gauge theories, despite the fact that instantons, sphalerons, etc., 
are quite general constructions inherent to many quantum field theories (QFT).
In this paper, we address some models {\em without gauge fields} in which degenerate ground states exist due to supersymmetry (as opposed to gauge-equivalent, topologically different ``pre-vacua'' in non-supersymmetric Yang-Mills theories).  We will limit ourselves to non-chiral 2D and 3D models with two supercharges (minimal SUSY, ${\mathcal N}=1$) or four supercharges (${\mathcal N}=2$ in two and three dimensions.)

In this work, we focus on two-dimensional models of the Wess-Zumino type which support kinks,
instantons and sphalerons with rather peculiar features in the fermion sector scarcely discussed in the literature.

The initial impetus for the present work was given by the recent publication by Evslin et al. \cite{Evslin:2025zcb}
in which a non-supersymmetric model (the so-called Montonen–Sarker–Trullinger–Bishop model (MSTB) \cite{Montonen:1976yk,Sarkar:1976vr} of real scalar fields was considered.
This model has two degenerate kinks which can mix due to an instanton. 

Similar systems have been described in the condensed matter literature.
Ref. \cite{Takagi:96a} % Eq. (14)-(16): oscillation frequency between two different Bloch walls in a 1D spin chain. Phase angle is a conjugate to the wall center.
discusses two different types of Bloch walls in a spin chain.
The tunneling probability between these walls is nonzero, which can lead to a corresponding observable resonant frequency.
Ref. \cite{DubeStamp1998} % Eq. (2.4) is nearly exactly the MSTB kink.
discusses a similar process of tunneling between two domain walls in a 1D material.
In Ref. \cite{ShibataTakagi2000}, the authors derive the effective action on the domain walls and compute the tunneling probability between misaligned walls.
Ref. \cite{GalkinaIvanov2008} studies further properties of the effective dynamics of these walls.
For more modern references, see, e.g., \cite{Izquierdo:2009ci,HongoFujimoriMisumiNittaSakai2020}.

Adding SUSY to this mix results in the possibility of having fermionic zero modes.
In this paper, we consider two examples of such models.
The first is the MSTB model supersymmetrized to $\mathcal{N}=(1,1)$ (which turns out to be nontrivial).
The second is a $\mathcal{N}=(2,2)$ supersymmetric model with a superpotential motivated by the Affleck-Dine-Seiberg (ADS) superpotential.

In both cases we first study the bosonic sector, finding the degenerate kinks, a sphaleron-like solution, and the instanton configuration.
If we restrict ourselves to the bosonic sector, the mixing of kinks does happen.
Restrictions posed by SUSY even on the bosonic sector allow for some (classically) exact analytical results.

As the next step, we take the fermionic sector into account. 
In the supersymmetric case we find that the fermionic zero modes localized on the instanton suppress the mixing of kinks. 
Thus, the kinks stay degenerate on the quantum level; otherwise, they could lose the BPS property.
Along the way we obtain some other new results, for example, a new way of supersymmetrizing the old $\varphi^4$ model.

We also note that in this paper we focus on the case when the multiplicity of the degenerate kinks is finite.
One may say that the kinks form a finite-dimensional multiplet.
There are other models where the kinks form degenerate families characterized by continuous moduli spaces (see, e.g., \cite{Shifman:1997wg,Gauntlett:2000ib,Tong:2002hi}),
but those are not directly related to the present work.

This paper is organized as follows. 
In Sec.~\ref{sec:N=11_generalities} we consider the generalities of the ${\mathcal N}=1$ theories of real scalar fields in two dimensions. 
In Sec.~\ref{sec:mstb} we study the MSTB model and its $\mathcal{N}=1$ supersymmetrization. The superpotential in this model turns out not to be smooth at one point, which causes a variety of interesting phenomena. We find that there are two regimes in this model, and in one regime it has two degenerate BPS kinks, a semi-BPS sphaleron, and a non-BPS instanton. On the quantum level, we find that supersymmetric vacua are lifted at some critical value of the coupling constant.
In Sec.~\ref{sec:ADS_motivated_model} we consider an $\mathcal{N}=2$ supersymmetric model that also has degenerate kinks.
The extended supersymmetry of this model allows for exact results concerning the instanton action and zero modes.
In Sec.~\ref{sec:concl} we briefly summarize the results and discuss the future directions of research.
Appendix~\ref{sec:phi4_family} presents a new family of supersymmetric models of the $\varphi^4$ type, all of which have the same ``old'' bosonic part.
Appendix~\ref{sec:more_on_central_charges} summarizes some facts about the central charges in 2D, while in Appendix~\ref{sec:almost_bps_2} we note one curiosity about the $\mathcal{N}=1$ MSTB model.

Some computations and plots for this paper were made with the help of \textit{Mathematica}.
The corresponding code is available at \cite{github-kink-inst}.

\section{Generalities of \boldmath{${\mathcal N}=1$} theories of real scalar fields in 2D}
\label{sec:N=11_generalities}

\subsection{Action and supercharges}

The corresponding Wess-Zumino model in two dimensions, with
two supercharges contains $k$ real bosonic fields $\phi ^a$ ($a=1,\ldots,k$)
and the same number of real (Majorana) two-component fermionic fields $\psi_\alpha^a$
($\alpha =1,2$). The model is characterized 
by a
superpotential ${\cal W}(\phi )$ which can be an
arbitrary function. We will assume that ${\cal W}(\phi )$ has more than one critical point, i.e. points where
$\partial_a{\cal W}(\phi )=0$.

The Lagrangian to be considered is 
\begin{equation}
\begin{aligned}
	{\cal L} = \!\frac{1}{2} &
	\left[\,\partial_\mu\phi^a \, 
	\partial^\mu\phi^b + \bar\psi^a \,i\gamma^\mu 
	\partial_\mu \psi^b +F^a F^b\right]
	\\
	&+\! F^a\partial_a {\cal W} - \frac1 2 \,
	(\partial_a \partial_b {\cal W})\,\bar\psi^a
	\psi^b
	\,,
\end{aligned}
\label{sigmaL}
\end{equation}
where 
$F^a$ is the auxiliary field, 
$$
F^a=-\partial^a  {\cal W}\,\, {\rm and}\,\,\partial_a = \frac{\partial }{\partial \phi^a } \,.
$$
The  \none
supersymmetry of the model generates two
supercharges,
\begin{equation}
	Q_\alpha=\int {\rm d} z \,J_{\alpha}^0\,,\quad \alpha=1,2\,,\quad 
	J^\mu_\alpha=\Big[\left(\not \!\!\,\partial \,\phi^a-i \,F^a\right)
	\gamma^\mu\Big]_{\alpha\beta}\,\psi^\beta_a \,,
	\label{supercur}
\end{equation}
where $J^\mu$ is the conserved supercurrent.

The \none SUSY
algebra\footnote{To emphasize its non-chiral nature it is often denoted as
	${\cal N}\!=\!\{1,1\}$.} corresponding to Eq. (\ref{sigmaL}) in two dimensions $\{t,z\}$ is as follows,
\begin{equation}
	\{ Q_\alpha\,,  {\bar Q}_\beta \}=2\left(\gamma^\mu 
	P_\mu + i \gamma^5 {\cal Z}\right)_{\alpha\beta}\, ,
	\qquad 
	\left[Q_\alpha, P_\mu\right]=\left[Q_\alpha, {\cal Z}\right]=0\,.
	\label{alg}
\end{equation}
Here $P_\mu $, ($\mu
=1,2$) is the energy-momentum vector  and  ${\cal Z}$ is the central charge, 
\beq
\Zc=\int  {\rm d} z\, \partial_z \phi^a 
\,\partial_a {\cal W}\,.
\label{cch}
\eeq
Moreover, 
${\bar Q}_\beta=Q_\alpha (\gamma^0)_{\alpha\beta}$.
For the $\gamma$ matrices in Minkowski spacetime, we follow the convention of \cite{Shifman:1998zy,Losev:2001uc,Losev:2000mm}, which is suitable for Majorana fermions:
\begin{equation}
	\gamma^0 \equiv \gamma^t = \sigma_2 \,, \quad
	\gamma^1 \equiv \gamma^z = i \sigma_3 \,, \quad
	\gamma^5 = \gamma^0 \gamma^1 = -  \sigma_1 \,.
\label{gamma_matrix_convention_MSTB}
\end{equation}
The Dirac conjugation and charge conjugation are given by
\begin{equation}
	\bar{\psi} = \psi^\dagger \gamma^0 \,, \quad
	\psi_{(c)} = \mathcal{C} \psi^{*} \,.
\end{equation}
Here, $\mathcal{C}$ is the charge conjugation matrix satisfying
\begin{equation}
	\mathcal{C} \left( \gamma^\mu \right)^{*} = - \gamma^\mu \mathcal{C} \,.
\label{charge_conjugation_matrix_condition_mink}
\end{equation}
Representation \eqref{gamma_matrix_convention_MSTB} is convenient because the gamma matrices are purely imaginary, and $\mathcal{C}$ becomes the identity matrix,
\begin{equation}
	\mathcal{C} = \mathbb{I}_{2 \times 2} \,.
\end{equation}
Majorana condition
\begin{equation}
	\psi_{(c)} = \psi
\label{Majorana_condition}
\end{equation}
in this representation simply becomes the condition of reality of both components of $\psi$.

Equation (\ref{sigmaL}) implies that the fermion mass matrix has the form
\beq
{\cal M}_{ab}= \partial_a \partial_b {\cal W}(\phi)\Big|_{\phi = \phi_{\rm vac}} \,,
\label{mm}
\eeq
where the vacuum values of the fields $\phi^a$ are determined by solving the set of equations
\beq
\partial_a  {\cal W} = 0\,, \,\,\, \forall a \,.
\eeq

The Dirac equation following from the Lagrangian \eqref{sigmaL} is
\begin{equation}
	i \gamma^\mu \partial_\mu \psi^a - (\partial_a \partial_b {\cal W}) \psi^b = 0 \,.
	\label{dirac_equation_1}
\end{equation}

The central charge is nonvanishing in the case of  solitons (kinks) 
provided that \cite{Witten:1978mh}
\begin{equation}
	{\cal Z}_0={\cal W}_f-{\cal W}_i\equiv
	{\cal W} \Big(\phi (z\to + \infty)\Big)-{\cal W} \Big(\phi (z\to 
	-\infty)\Big)\neq 0\,.
\label{zclas}
\end{equation}
The subscripts $f$ and $i$ stand for final and initial, respectively. In what follows we will sometimes write $\Delta{\cal W}\equiv {\cal W}_f-{\cal W}_i$.
Our convention is as follows: if  $\mathcal{Z}\!>\!0$ we deal with a kink; otherwise (i.e. ${\cal Z}\!<\!0$) we deal with an anti-kink. The subscript $0$ in Eq. (\ref{zclas}) warns the reader that
the central charge on the left-hand side is ``bare.''
As reviewed below, in the \none problems discussed in this section the superpotential is not protected against
perturbative corrections \cite{Losev:2001uc,Losev:2000mm,Shifman:1998zy}.

\subsection{2D specific anomalies}

Two anomalies specific to 2D field theories have been studied previously \cite{Shifman:1998zy,Losev:2001uc}.
In this section we briefly review them since they play a role in our present consideration.

\subsubsection{Loss of fermion parity in topologically nontrivial sectors}
\label{sec:F_loss}

In Sec.~\ref{sec:mstb} we consider $\mathcal{N}=(1,1)$ supersymmetrization of the MSTB model. 
This model is an example of a more general class of models with a non-perturbative fermion parity anomaly, as we now review.

The fermion sector consists of two Majorana fields $\psi_a$ ($a=1,2$ is the flavor index), each of them is a two-component real spinor. 
One feature of this model is that there is no $U(1)$ symmetry of the fermionic fields, and the ``fermion number'' cannot be defined.
However, one can still define the fermion parity $(-1)^F$. 

In the topologically trivial sector, in any diagram the $\psi_a$ field line is characterized by  $(-1)^F=-1$, while for the corresponding bosons $\phi_a$ we have $(-1)^F=1$. 
The fermion parity is conserved. 
The irreducible supermultiplets are two-dimensional, built from the quanta of $\phi_a$ and $\psi_a$ fields.

However, in passing to topologically nontrivial sectors one loses the fermion parity for BPS-saturated kinks.
The supermultiplet of such a kink is one-dimensional,
i.e. the BPS kink satisfies the following equation,
\beq
Q_1|{\rm BPS\,kink}\rangle =\sqrt{2\mathcal Z}|{\rm BPS\,kink}\rangle \,,
\label{non-reduce}
\eeq
where $Q_1$ is the broken supercharge (for the conserved supercharge $Q_2$ we have \,\,$Q_2|{\rm BPS\,kink}\rangle=0$) and  $\mathcal Z$ is the central
charge. 
As seen from Eq. (\ref{non-reduce}) the supermultiplet containing the BPS-saturated kink is non-reducible. The fermion parities of the supercharges 
which normally would  be $-1$ are not defined. This phenomenon is due to the fact that the BPS kink has only one $\psi$ field zero mode. This is true both in the model studied in Ref. \cite{Losev:2001uc} and in the model under consideration in this paper, see Sec.~\ref{sec:ferm_kink_MSTB}.

\subsubsection{ Anomaly in the superpotential}
\label{sec:W_anomaly_general}

The anomaly in ${\mathcal N}=(1,1)$ theories was first discovered in a model with a single real boson field (and its fermion superpartner) and the sine-Gordon model almost 30 years ago.
Depending on a specific regularization method used, computations can yield the kink mass which may or may not violate the BPS bound, see e.g. \cite{Rebhan:1997iv,Nastase:1998sy,Graham:1998qq}.
This issue was later clarified in \cite{Shifman:1998zy}, where supersymmetry-preserving regularization was used.

Here we will outline the derivation of \cite{Shifman:1998zy} assuming that the number of fields is arbitrary. 
We start from the Lagrangian \eqref{sigmaL} implying the classical expression
for the supercurrent quoted in \eqref{supercur}. 
Classically,
\beq
\Big(\gamma^\mu J_\mu \Big)_{\rm class} =-2i F^a\psi_a
\label{gammaJ}
\eeq 
due to the fact that $\gamma_\mu\gamma^\nu\gamma^\mu=0$ in two dimensions. As we will see, the right-hand side of (\ref{gammaJ}) changes upon the UV regularization. The simplest one is the dimensional regularization%
\footnote{Additional verification is provided by a regularization through higher derivatives \cite{Shifman:1998zy} and dimensional reduction \cite{Rebhan:2002yw}. } 
to be used below in which  $\gamma_\mu\gamma^\nu\gamma^\mu=D-2$. 

To carry out the one-loop calculation\footnote{Why one loop is sufficient will be explained below.}, we apply the background field technique; namely we substitute the field $\phi^a$
by its background and quantum parts, $\phi^a$ and $\chi^a$, respectively,
\begin{equation}
	\phi^a \longrightarrow \phi^a+\chi^a \,.
\label{backsubst}
\end{equation}
In this way,  we arrive at 
\begin{equation}
	\begin{aligned}
		\left( \gamma^\mu J_\mu\right)_{\rm anom} 
		&= (D-2)\,  (\partial_\nu\phi^a )\, \gamma^\nu\psi_a\to	-(D-2)\,  \chi^a  \gamma^\nu\partial_\nu\psi_a \\
		&=	i\,(D-2)\, \chi^a \, \Big(\partial_a\partial_b{\cal W}(\phi+\chi)\Big)\,\psi^b	\,, \\
		&\to i\,(D-2)\,\langle 0|\chi^a\chi^c|0\rangle\, \Big(\partial_a\partial_b\partial_c{\cal W}(\phi)\Big) \psi^b \,,
    \end{aligned}
\end{equation}
where an integration by parts has been carried out, and a total
derivative term  is omitted (on dimensional grounds it vanishes in
the limit $D=2$). We also used  the equation of motion for  the $\psi^a$
field. Moreover,
\beq
\langle 0|\chi^a\chi^c|0\rangle =\delta_{ab}\, i\int \!\frac{{\rm d}^D p}{(2\pi)^D} \, \frac{1}{p^2-m^2} \;
\eeq
and
\beq
\left( \gamma^\mu J_\mu\right)_{\rm anom} =\frac{i}{2\pi} \Big(\partial_a\partial_b\partial^a{\cal W}(\phi)\Big)\,\psi^b \,.
\eeq

The next steps are quite straightforward. First, applying the canonical commutation relations we
arrive at the general formula
\begin{equation}
	\left\{J_{\alpha}^\mu, \bar Q_\beta \right\}
	=2\,(\!\gamma_\nu)_{\alpha\beta} \,\vartheta^{\mu\nu} +
	2i\,(\!\gamma^5)_{\alpha\beta}\, \zeta^\mu 
\label{celoc}
\end{equation}
telling us that the conserved supercurrent and the energy-momentum tensor $\vartheta^{\mu\nu}$
are in one and the same supermultiplet. Here
$\zeta^\mu$ is the conserved topological current,
\begin{equation}
	\zeta^\mu=\epsilon^{\mu\nu} \partial_\nu {\cal W}\,.
\label{topcurr}
\end{equation}
Multiplying \mbox{Eq.\ (\ref{celoc})} by $\gamma_\mu$ from the
left, we get the supertransformation of $\gamma_\mu J^\mu$,
\beq
\frac{1}{2} \, \left\{ \gamma^\mu J_\mu\,, \bar Q
\right\} = \vartheta^\mu_\mu +i \gamma_\mu \gamma^5
\,\zeta^\mu
\, .
\label{clgscst}
\eeq
This equation  establishes the supersymmetric relation between
$\gamma^\mu J_\mu$, $\vartheta^\mu_\mu $ and $\zeta^\mu$. It   remains \textit{valid with quantum corrections included}. 

At the same time, individual terms in (\ref{clgscst}) could be changed.
Classically, the trace of the energy-momentum tensor is given by
\begin{equation}
	\left(\vartheta^\mu_{\,\mu}\right)_{\rm class}=F^aF_a + \frac 12\,(\partial_a\partial_b{\cal
		W})\, \bar\psi^a\psi ^b \,.
\end{equation}

Next, we observe that the  supertransformation of the anomalous term in
$\gamma^\mu J_\mu$ is
\begin{equation}
	\begin{aligned}
		\frac{1}{2} \, \left\{ \left(\gamma^\mu	J_\mu \right)_{\rm anom},\bar Q	\right\} 
		&= \left\{\frac{1}{8\pi}\, \Big(\partial_b\partial_c\partial_a\partial^a{\cal W}(\phi)\Big)\bar\psi^b\psi^c	-\frac{1}{4\pi}\Big(\partial_b\partial_a\partial^a{\cal W}(\phi)\Big)F^b\right\} \\
		&+ i \gamma_\mu \gamma^5 \epsilon^{\mu\nu}\partial_\nu	\left(\frac{1}{4\pi}\partial_a\partial^a{\cal W}\right)\, .
	\end{aligned}
\end{equation}
The term in the braces on the right-hand side is the anomaly in the trace of
the energy-momentum tensor, the second term
represents the anomaly in the topological current.  After the inclusion of the anomaly term the topological current takes the form,
\beq
\zeta^\mu = \epsilon^{\,\mu\nu}\partial_\nu \left( {\cal W}+
\frac{1}{4\pi}\,
\partial_a \partial^a  {\cal W}\right) \,.
\label{ccdanoc}
\eeq
This gives the anomalous shift in the superpotential
\beq
\Wc(\phi ) \longrightarrow \widetilde{\Wc}(\phi) \equiv\Wc (\phi)+\frac{1}{4\pi}\pt^a\pt_a\Wc (\phi)\,.
\label{sevenan}
\eeq
Therefore, the central charge in (\ref{zclas}) is 
to be replaced by
\beq
{\cal Z}_0 \longrightarrow \Zc = \widetilde{\cal W} \Big(\phi (z\to + \infty)\Big)-\widetilde{\cal W} \Big(\phi (z\to 
-\infty)\Big)\,.
\label{eightan}
\eeq
It is worth emphasizing that the second (anomalous) term in the right-hand side of (\ref{sevenan}) is the {\em only} contribution coming from the UV. It comes from the first loop. 
Higher loops are well defined in the UV and do not contribute to the anomaly under consideration \cite{Shifman:1998zy}. 
The replacement (\ref{sevenan}) must be performed wherever the bare superpotential appears.

This anomaly will be computed for the MSTB model in Sec.~\ref{sec:W_anom} below.

\section{ \boldmath MSTB model with $\mathcal{N}=(1,1)$ supersymmetry}
\label{sec:mstb}

As was mentioned above, the bosonic MSTB model was introduced in \cite{Montonen:1976yk,Sarkar:1976vr}.
This is a (1+1)-dimensional model of two real scalar fields with global O(2) symmetry that is broken both spontaneously and explicitly.
An interesting feature of this model is the presence of two vacua and two non-equivalent kinks interpolating between those vacua.
For a historical overview with references, as well as a comprehensive account of the kink dynamics in the bosonic case, see \cite{Alonso-Izquierdo:2018uuj}.

A recent paper \cite{Evslin:2025zcb} suggests that, in the bosonic setup, these kinks in fact mix quantum mechanically due to an instanton effect.
In this section we will study this phenomenon in the supersymmetrized version of this model.
Along the way we will review (and sometimes re-derive by a different method) some properties of the bosonic kinks and the instanton interpolating between them.
In particular, we will provide a shortcut for (approximately) computing the instanton action in subsection~\ref{sec:instanton_MSTB}.

\subsection{Bosonic setup}

Let us start by recalling the formulation of the original MSTB model.
We introduce two real scalar fields $\phi_a$, where $a=1,2$ is a flavor index.
The action in Minkowski spacetime is given by
\begin{equation}
	S = \int dzdt \, \left\{ \frac{1}{2} (\partial_\mu \phi_a) (\partial^\mu \phi_a) - U(\phi_1, \phi_2) \right\} \,.
\label{MSTB_action}
\end{equation}
Here and below, summation over repeated flavor indices $a,b$ and/or over the repeated spacetime indices $\mu,\nu$ is implied, unless stated otherwise.
$U(\phi_1, \phi_2)$ is the potential, see Fig.~\ref{fig:MSTB_potential_V}:
\begin{equation}
	U(\phi_1, \phi_2) = \frac{\lambda^2}{2} \left(\phi_1^2 + \phi_2^2 - \frac{m^2}{4 \lambda^2} \right)^2  + \frac{ \kappa^2 m^2 }{8} \phi_2^2 \,.
\label{MSTB_potential}
\end{equation}
Here, $\lambda,m > 0$ and $\kappa \in \mathbb{R}$ are constants.
We have chosen the notation for the various parameters as most convenient in the supersymmetric setup\footnote{\label{ft:notation} Our notation is related to the notation of \cite{Evslin:2025zcb} as follows: $m_\text{\cite{Evslin:2025zcb}} = m_\text{our}$, $\lambda_\text{\cite{Evslin:2025zcb}} = 2\lambda_\text{our}^2$, $(1-\alpha^2)_\text{\cite{Evslin:2025zcb}} = \kappa^2_\text{our}$. The parameter $\alpha$ defined by the latter relation is sometimes used in this work as well, in particular in Sec.~\ref{sec:worldline_inst}.}.

Let us briefly discuss the global symmetries of this model.
At a generic $\kappa$, there are two internal $\mathbb{Z}_2$ symmetries which we will call $S$ and $C$, acting as
\begin{subequations}
	\begin{equation}
		S: \quad \phi_1 \to - \phi_1 \,;
		\label{symmetries_mstb_S}
	\end{equation}
	\begin{equation}
		C: \quad \phi_2 \to - \phi_2  \,.
		\label{symmetries_mstb_C}
	\end{equation}
\label{symmetries_mstb}
\end{subequations}
The $S$ symmetry is however spontaneously broken, as the minima of the classical potential \eqref{MSTB_potential} are given by
\begin{equation}
	\text{Vac}_\pm \, : \quad
	\phi_1 = \pm \frac{m}{2 \lambda} \,, \quad
	\phi_2 = 0 \,.
	\label{MSTB_vacua}
\end{equation}
As we will see later, the symmetry $S$ is going to be explicitly broken by the supersymmetric coupling to fermions, while $C$ is spontaneously broken in the kink background.
In the special case $\kappa = 0$, the $\mathbb{Z}_2 \times \mathbb{Z}_2$ is enhanced to $O(2)$, and the vacuum manifold becomes a circle.

\subsection{\boldmath $\mathcal{N}=(1,1)$ supersymmetrization}
\label{sec:MSTB_supersymmetrization}

\begin{figure}[t]
	\centering
	\begin{subfigure}[b]{0.4\textwidth}
		\centering
		\includegraphics[width=\textwidth]{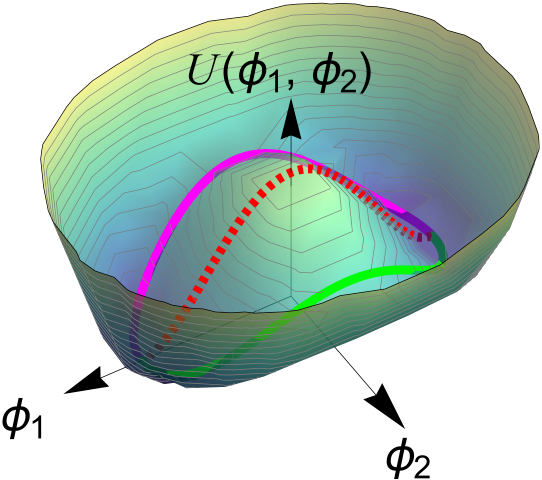}
		\subcaption{$U(\phi_1,\phi_2)$}
		\label{fig:MSTB_potential_V}
	\end{subfigure}
	
	\begin{subfigure}[b]{0.4\textwidth}
		\centering
		\includegraphics[width=\textwidth]{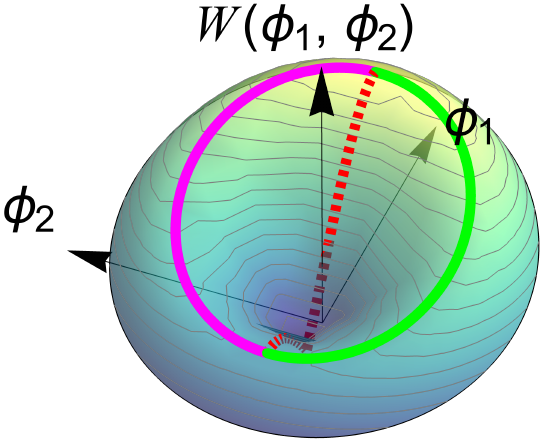}
		\subcaption{$\mathcal{W}(\phi_1,\phi_2)$}
		\label{fig:MSTB_potential_W1}
	\end{subfigure}
	~
	\begin{subfigure}[b]{0.45\textwidth}
		\centering
		\includegraphics[width=\textwidth]{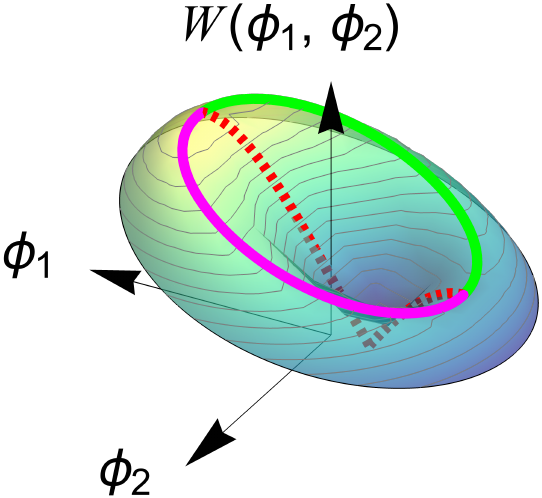}
		\subcaption{$\mathcal{W}(\phi_1,\phi_2)$}
		\label{fig:MSTB_potential_W2}
	\end{subfigure}
	\caption{\small
		(Super)Potentials for the MSTB model for $\lambda = m = 1$.
		(\subref{fig:MSTB_potential_V}) Potential \eqref{MSTB_potential} at $\kappa=0.8$.
		(\subref{fig:MSTB_potential_W1}) and (\subref{fig:MSTB_potential_W2}) Superpotential \eqref{MSTB_superpotential} at $\kappa = 0.3$ (the same plot from two different points of view).
		The values of $\kappa$ are chosen so that the characteristic features are most prominent visually.
        The thin gray contour lines mark the height of $U$ or $\mathcal{W}$.
		The profiles of the BPS degenerate kinks \eqref{MSTB_kinks} are shown by the thick solid lines, magenta and green respectively.
		The sphaleron profile \eqref{MSTB_sphaleron} is shown by the thick dashed red line.
		On (\subref{fig:MSTB_potential_W1}) and (\subref{fig:MSTB_potential_W2}), the BPS (and semi-BPS) profiles --- the thick lines --- can be seen as trajectories of viscous honey trickling down on the superpotential landscape (different trajectories correspond to different infinitesimal initial speeds).
	}
	\label{fig:MSTB_potential}
\end{figure}

As reviewed in Sec.~\ref{sec:N=11_generalities}, each real supermultiplet consists of a real scalar field $\phi_a$ and a corresponding Majorana fermion $\psi^a$.
The bosonic part of the supersymmetrized action should still reproduce \eqref{MSTB_action}.
Thus, the first step is to find a suitable superpotential $\mathcal{W}(\phi_1, \phi_2)$ such that it reproduces the potential \eqref{MSTB_potential},
\begin{equation}
	U(\phi_1, \phi_2) = \frac{1}{2} \left( \pdv{ \mathcal{W} }{ \phi_a } \right)^2 \,.
\label{U_from_W_N11}
\end{equation}
Generically, finding such a superpotential in a theory with multiple flavors can be non-trivial.
Fortunately, in the present case one can guess it by recalling a well-known example of a theory with a single real superfield $\varphi$ and superpotential
\begin{equation}
	\mathcal{W}^{\text{toy}} = \frac{m^2}{ 4\lambda} \varphi - \frac{\lambda}{3} \varphi^3 \,.
\end{equation}
The scalar potential in that simple model is 
\begin{equation}
	U^{\text{toy}} = \frac{1}{2} \left( \pdv{ \mathcal{W}^{\text{toy}} }{ \varphi } \right)^2 
	= \frac{\lambda^2}{2} \left( \varphi^2 - \frac{m^2}{4 \lambda^2} \right)^2 \,.
\end{equation}
Based on this example, one can cook up the superpotential for the MSTB model at hand.
It turns out to be given by:
\begin{equation}
	\mathcal{W}(\phi_1, \phi_2) = \frac{m^2}{ 4\lambda} \Upsilon - \frac{\lambda}{3} \Upsilon^3 + \frac{1}{2} \kappa m \phi_1 \Upsilon \,, \quad
	\Upsilon \equiv \sqrt{ \left(\phi_1 + \kappa \frac{m}{2 \lambda}  \right)^2 + \phi_2^2 } \,.
\label{MSTB_superpotential}
\end{equation}
The square root here is understood to be non-negative.
This superpotential, as a function of the real fields $\phi_1$ and $\phi_2$, is smooth everywhere in the field space, except for the point $\Upsilon=0$, where
\begin{equation}
	\phi_1 = - \kappa \frac{m}{2 \lambda} \,, \quad
	\phi_2 = 0 \,.
\label{W_singularity_1}
\end{equation}
In the vicinity of this point the superpotential is given by
\begin{equation}
	\mathcal{W} = \frac{m^2 (1 - \kappa^2)}{ 4\lambda} \Upsilon + O(1) \,, \quad
	\Upsilon \to 0 \,.
\label{W_singularity_2}
\end{equation}
For a visual representation, see Fig.~\ref{fig:MSTB_potential}.

%%%%%%%%%%%%%%%%%%%%%%%%%%%%%%%%%%%%%%%%%%%%%%%%%%%%%%%%%%
%%%%%%%%%%%%%%%%%%%%%%%%%%%%%%%%%%%%%%%%%%%%%%%%%%%%%%%%%%
%%%%%%%%%%%%%%%%%%%%%%%%%%%%%%%%%%%%%%%%%%%%%%%%%%%%%%%%%%
%%%%%%%%%%%%%%%%%%%%%%%%%%%%%%%%%%%%%%%%%%%%%%%%%%%%%%%%%%
%\clearpage
\FloatBarrier

The supersymmetric coupling to fermions following from this superpotential, see Eq.~\eqref{sigmaL}, turns out to violate the $\mathbb{Z}_2$ symmetry $S: \ \phi_1 \to - \phi_1$ that we had on the bosonic level, see Eq.~\eqref{symmetries_mstb_S}.
As long as the parameter $\kappa$ is non-vanishing, the superpotential \eqref{MSTB_superpotential} (as well as its second derivatives) strictly speaking does not have simple transformation properties under the flip of $\phi_1$.
This symmetry is recovered only in the limit $|\phi_1| \gg |\kappa| m / \lambda$, for example, when $\kappa = 0$, or when we are far enough from the origin of the field space, or if we supplement it by a sign flip for $\kappa$ as well.

The other symmetry $C: \ \phi_2 \to - \phi_2 $ survives when supplemented by the corresponding fermion transformation $\psi_2 \to - \psi_2$.

Vacua of the theory are given by the critical points $\partial_a \mathcal{W} = 0$; of course, the solutions of this equation coincide with \eqref{MSTB_vacua} for any value of $\kappa$.
Values of the superpotential \eqref{MSTB_superpotential} on the vacua \eqref{MSTB_vacua} are given by
\begin{equation}
	\begin{aligned}
		\mathcal{W} \Big|_{ \text{Vac}_+ } &= \frac{ m^3 }{ 24 \lambda^2 }  |1-\kappa| (1 - \kappa) (2 + \kappa) \,, \\
		\mathcal{W} \Big|_{ \text{Vac}_- } &= \frac{ m^3 }{ 24 \lambda^2 }  |1+\kappa| (1 + \kappa) (2 - \kappa) \,. \\
	\end{aligned}
\label{MSTB_vacua_Wval}
\end{equation}

These expressions, Eq.~\eqref{MSTB_vacua_Wval}, hint that something non-trivial happens at $\kappa = \pm 1$.
And indeed, it is known \cite{Montonen:1976yk} that for $|\kappa|<1$ this model enjoys two stable degenerate kinks and one unstable kink (sphaleron).
At $\kappa = \pm 1$ these three solutions merge, and for $|\kappa| > 1$ there is only one stable kink ($\kappa = 2$ is also an interesting point, see Appendix~\ref{sec:almost_bps_2}).
Below we will briefly review these bosonic solutions, and for each of those we are going to discuss the fermionic sector.

\subsection{ \boldmath Two degenerate kinks at $|\kappa| < 1$}

For $|\kappa| < 1$, the model at hand enjoys two kink solutions with classically degenerate energies.
These kinks interpolate between the same pair of vacua, and so their topological charges are equal.

\subsubsection{Kink profiles}

The explicit form of the superpotential allows us to easily write down Bogomolny representation for the energy of static solutions
\begin{equation}
	\begin{aligned}
		E 
		&= \int dz \left[ \frac{1}{2} (\partial_z \phi_a)^2 + U(\phi_1, \phi_2) \right] \\
		&= \int dz \sum_a \frac{1}{2} \left[ \partial_z \phi_a \pm \pdv{ \mathcal{W} }{ \phi_a } \right]^2 + |\mathcal{Z}_\text{kink}| \,,
	\end{aligned}
\label{MSTB_energy_bogomolny}
\end{equation}
where, classically, the central charge $\mathcal{Z}_\text{kink}$ is given by
\begin{equation}
	\begin{aligned}
		\mathcal{Z}_\text{kink}
		&= \int dz \left[ \pdv{ \mathcal{W} }{ \phi_a } \, \partial_z \phi_a  \right]
		&= \mathcal{W} \Big|_{ \text{Vac}_+ } - \mathcal{W} \Big|_{ \text{Vac}_- } 
	\end{aligned}
\label{MSTB_central_charge}
\end{equation}
for a classical trajectory interpolating from vacuum $\text{Vac}_-$ at $z \to - \infty$ to the vacuum $\text{Vac}_+$ at $z \to + \infty$.
We call such a trajectory ``kink'', while the trajectory going in the opposite direction will be referred to as ``antikink'' (this naming convention coincides with \cite{Alonso-Izquierdo:2018uuj}).

\begin{figure}[t]
	\centering
	\includegraphics[width=0.5\textwidth]{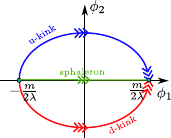}
	\caption{\small
		Field space spanned by $\phi_a$, $a=1,2$.
		Vacua \eqref{MSTB_vacua} are shown by two cyan dots.
		Kinks \eqref{MSTB_kinks} are shown by the blue (upper) and red (lower) solid lines respectively (see also Eq.~\eqref{ellipse_trajectory}).
		The sphaleron \eqref{MSTB_sphaleron} is shown by green line going through the center.
	}
	\label{fig:MSTB_kinks}
\end{figure}

From the representation \eqref{MSTB_energy_bogomolny} one can read off the first order equations
\begin{equation}
	\partial_z \phi_a = \pm \pdv{ \mathcal{W} }{ \phi_a } \qquad
	\text{(``$+$'' sign for kinks)} \,,
\label{MSTB_first_order_eq}
\end{equation}
and the corresponding rest energy $M_\text{kink} = | \mathcal{Z} |$.
The kink profiles are given by
\begin{equation}
	\phi_1^\text{kink}(z) = \frac{m}{2 \lambda} \tanh( \kappa m z / 2 ) \,, \quad
	\phi_2^\text{kink}(z) = \pm \frac{m}{2 \lambda} \sqrt{ 1 - \kappa^2 } \frac{1}{ \cosh( \kappa m z / 2 ) } \,.
\label{MSTB_kinks}
\end{equation}
Note that these two kink trajectories live on the ellipse
\begin{equation}
	\phi_1^2+ \frac{\phi_2^2}{  1 - \kappa^2 } = \left( \frac{m}{2\lambda} \right)^2 \,.
\label{ellipse_trajectory}
\end{equation}
Thus, $\kappa$ plays the role of the ellipse's eccentricity.
Below we will sometimes refer to the kink in the upper half-plane (``$+$'' sign in Eq.~\eqref{MSTB_kinks}) as a ``u-kink'', while the kink in the lower half-plane (``$-$'' sign in Eq.~\eqref{MSTB_kinks}) may be referred to as a ``d-kink''.
These kinks are exchanged by the $\mathbb{Z}_2$ transformation \eqref{symmetries_mstb_C}.
For a visual representation, see Fig.~\ref{fig:MSTB_kinks} and also Figs.~\ref{fig:MSTB_potential}, \ref{fig:ferm_eig}.

Rest energies of the kinks are easily determined from the Bogomolny representation \eqref{MSTB_energy_bogomolny}.
They are given by the absolute value of the central charge \eqref{MSTB_central_charge},
\begin{equation}
	M_\text{kink} 
	= \abs{ \mathcal{W} (\text{Vac}_+ ) - \mathcal{W} ( \text{Vac}_- ) }
	= \frac{  m^3 }{ 12 \lambda^2 } |\kappa| (3 - \kappa^2) \,.
\label{MSTB_kink_mass}
\end{equation}
We emphasize that these kinks exist only when $\abs{ \kappa } < 1 $.
At the critical values of $\kappa$ maximizing the right-hand side of (\ref{MSTB_kink_mass}), i.e. $\kappa= \pm 1$,  the kink mass \eqref{MSTB_kink_mass} becomes degenerate with the sphaleron mass \eqref{MSTB_sphaleron_mass} below.

\subsubsection{Bosonic modes}

The equation for modes around the kink solution is
\begin{equation}
	\left[ \partial_t^2 - \partial_z^2 \right] \delta\phi_a + [\partial_a \partial_b U(\phi_1,\phi_2) ] \delta\phi_b = 0 \,,
\label{bosonic_mode_equation_dt}
\end{equation}
or, equivalently,
\begin{equation}
	- \partial_z^2  \delta\phi_a + [\partial_a \partial_b U(\phi_1,\phi_2) ] \delta\phi_b = \omega^2 \delta\phi_a \,
\label{bosonic_mode_equation_omega}
\end{equation}
where $\omega^2$ are eigenvalues. Moreover, $[\partial_a \partial_b U(\phi_1,\phi_2) ]$ should be evaluated on the kink background.
The kink \eqref{MSTB_kinks} has a single zero mode $\sim \partial_z \phi_a^\text{kink}(z)$:
\begin{equation}
\begin{aligned}
	\delta\phi_{a=1}(z) &= \frac{ 1 }{ \cosh[2]( \kappa \frac{mz}{2} ) } \sim (\phi_{a=1})' \,, \quad \\[2mm]
	\delta\phi_{a=2}(z) &= \mp \frac{ \sqrt{1-\kappa^2} \sinh( \kappa \frac{mz}{2} ) }{ \cosh[2]( \kappa \frac{mz}{2} ) } \sim (\phi_{a=2})'   \,,
\end{aligned}
\label{zero_mode_kink_bosonic}
\end{equation}
and a tower of modes with $\omega^2 > 0$.

\subsubsection{Fermionic modes}
\label{sec:ferm_kink_MSTB}

The fermion mass matrix, following from the superpotential, is given by 
\begin{equation}
	\mathcal{M}_{ab} = \partial_a \partial_b \mathcal{W} \,.
\label{mm_1}
\end{equation}
cf. \eqref{mm}.
Strictly speaking, Eq. (\ref{mm_1}) can be called ``the mass matrix'' only at the vacua, see (\ref{pmap}); for brevity we will use the same terminology off the vacuum point (\ref{MSTB_vacua}).

At the vacua \eqref{MSTB_vacua}, we have
\begin{equation}
	\mathcal{M}_{ab} \Big|_{ \text{Vac}_\pm }  = \begin{pmatrix}
		-m & 0 \\
		0 & \mp \frac{1}{2} \kappa m
	\end{pmatrix} \,.
\label{pmap}
\end{equation}
Therefore, as long as $\kappa \neq 0$ (which we assume here), one of the eigenvalues changes its sign in passing from one vacuum to another.
In other words, the relative Morse index of $\mathcal{W}$ is equal to one%
\footnote{\label{ft:ferm_mass_zero} Note that, since the matrix $\mathcal{M}_{ab}$ is real and symmetric, its eigenvalues are real. Therefore, in the present case one eigenvalue actually passes through zero. The corresponding localized massless fermionic states are known in condensed matter physics as Volkov-Pankratov states. }%
.
This means that any field configuration interpolating between these two vacua will have a fermionic zero mode.

For the kink \eqref{MSTB_kinks}, this fermionic zero mode is again proportional to $\partial_z \phi_a^\text{kink}(z)$.
It is generated by the action of one supercharge on the bosonic kink solution \eqref{MSTB_kinks}.
In the basis \eqref{gamma_matrix_convention_MSTB}, the zero mode is
\begin{equation}
	\begin{pmatrix} \psi_1^a \\ \psi_2^a \end{pmatrix}
	=
	\delta\phi_a(z) \begin{pmatrix} 0 \\ 1 \end{pmatrix}
	\,,
\label{zero_mode_kink_fermionic}
\end{equation}
where the profiles $\delta\phi_a(z)$ coincide with the profiles for the bosonic zero mode \eqref{zero_mode_kink_bosonic}.
Here, the lower index of the fermion $\psi_\alpha^a$ is the spinor index, $\alpha = 1,2$.

Let us now derive the Dirac equation in the kink background and check that the zero mode \eqref{zero_mode_kink_fermionic} indeed satisfies this equation.
Later we will use the Dirac equation to study (non-)zero modes in other backgrounds.

To this end we introduce two operators 
\begin{equation}
	\mathcal{D}_{ab} = \delta_{ab} \partial_z - [\partial_a \partial_b \mathcal{W}] \,, \quad
	\mathcal{D}_{ab}^\dagger = - \delta_{ab} \partial_z - [\partial_a \partial_b \mathcal{W}] \,,
\end{equation}
where $[\partial_a \partial_b \mathcal{W}] \equiv \mathcal{M}_{ab} $ is evaluated on a given bosonic background (e.g. kink or instanton).
Note that the conjugation $(\dagger)$ does not affect the flavor indices $a,b$.
In the gamma matrix representation \eqref{gamma_matrix_convention_MSTB}, the fermion equation \eqref{dirac_equation_1} can then be written explicitly as
\begin{equation}
	\begin{aligned}
		\mathcal{D}_{ab}^\dagger \psi_1^b + \partial_t \psi_2^a &= 0 \,, \\[2mm]
		- \partial_t \psi_1^a + \mathcal{D}_{ab} \psi_2^b &= 0 \,.
	\end{aligned}
\label{fermion_first_order_equations}
\end{equation}
In a static background $[\mathcal{D}_{ab}, \partial_t] =0$, and by taking extra derivatives and reshuffling these equations we can arrive at
\begin{equation}
	\begin{aligned}
		\partial_t^2 \psi_1^a + [\mathcal{D} \mathcal{D}^\dagger]_{ab} \psi_1^b &= 0 \,, \quad [\mathcal{D} \mathcal{D}^\dagger]_{ab} = - \delta_{ab} \partial_z^2 + \mathcal{M}_{ab'} \mathcal{M}_{b'b} - (\partial_z \mathcal{M}_{ab}) 
		\,; \\[3mm]
		\partial_t^2 \psi_2^a + [\mathcal{D}^\dagger \mathcal{D}]_{ab} \psi_2^b &= 0 \,, \quad [\mathcal{D}^\dagger \mathcal{D}]_{ab} = - \delta_{ab} \partial_z^2 + \mathcal{M}_{ab'} \mathcal{M}_{b'b} + (\partial_z \mathcal{M}_{ab}) 
		\,.
	\end{aligned}
\label{fermion_second_order_equations}
\end{equation}
In particular, if we take the background of a BPS kink satisfying the first order equations \eqref{MSTB_first_order_eq}, we obtain
\begin{equation}
	[\mathcal{D}^\dagger \mathcal{D}]_{ab} = - \partial_z^2 + [\partial_a \partial_b U(\phi_1,\phi_2) ] \,,
\label{DdagD}
\end{equation}
where $U$ is the scalar potential \eqref{U_from_W_N11}.

One can see that the equation for $\psi_2$ coincides with the equation for the bosonic modes \eqref{bosonic_mode_equation_dt}. 
Thus,  the equation for $\psi_2$ is the  partner of \eqref{bosonic_mode_equation_dt} in the sense of supersymmetric quantum mechanics.
In the zero mode sector this implies that for each bosonic zero mode $\delta\phi_a$ we have a corresponding fermionic mode,
\begin{equation}
	\psi^a = \delta\phi_a(z) \begin{pmatrix} 0 \\ 1 \end{pmatrix} \,,
\end{equation}
thus verifying \eqref{zero_mode_kink_fermionic}.
Generically, the partner operator $\mathcal{D} \mathcal{D}^\dagger$ does not have zero modes (even if it has accidental zero modes, they are not protected against small perturbations).

For non-zero modes, the standard lore is that for every bosonic mode with frequency $\omega$, i.e. a mode with
\begin{equation}
	- \partial_t^2 \delta\phi_a(z,t) = [\mathcal{D}^\dagger \mathcal{D}]_{ab} \delta\phi_b(z,t) = \omega^2 \delta\phi_a(z,t) \,,
\label{bosonic_mode_equation}
\end{equation}
there is a corresponding fermionic mode.
Explicitly, for the stationary mode
\begin{equation}
	\delta\phi_a(z,t) = \delta\phi_a(z) \cdot [ C_1 \sin(\omega t) + C_2 \cos(\omega t) ]
\end{equation}
with arbitrary real coefficients $C_1$ and $C_2$, we have the fermionic mode
\begin{equation}
	\begin{pmatrix} \psi_1 \\ \psi_2 \end{pmatrix} = 
	C_1 \begin{pmatrix} - \frac{1}{\omega} \mathcal{D}_{ab} \delta\phi_b(z) \cdot \cos(\omega t) \\ \delta\phi_a(z) \cdot \sin(\omega t) \end{pmatrix}
	+ C_2 \begin{pmatrix} \frac{1}{\omega} \mathcal{D}_{ab} \delta\phi_b(z) \cdot \sin(\omega t) \\ \delta\phi_a(z) \cdot \cos(\omega t) \end{pmatrix}
	\,.
\label{nonzero_fermionic_from_bosonic_modes}
\end{equation}
One can check this e.g. by plugging \eqref{nonzero_fermionic_from_bosonic_modes} into the Dirac equations \eqref{fermion_first_order_equations} and using \eqref{bosonic_mode_equation}.
This ensures, for example, that in the kink's background, the determinant for each Majorana fermion cancels the square root of the determinant for each corresponding real scalar, see e.g. \cite{tongQM}.

While it is true that this fermion satisfies equations \eqref{fermion_first_order_equations} and \eqref{fermion_second_order_equations} at a generic spacetime point, it might have isolated singularities that make this solution non-normalizable.
This might happen if derivatives of the superpotential contain a singularity; we will encounter this later.
The superpotential \eqref{MSTB_superpotential} is smooth on the kink trajectories \eqref{ellipse_trajectory}, as they do not pass through the irregular point \eqref{W_singularity_1}.

\subsection{Sphaleron}
\label{sec:sphaleron_mstb}

In the MSTB model at hand there is another topological solution present for all $\kappa$.
This solution corresponds to setting $\phi_2 \equiv 0$.
The potential \eqref{MSTB_potential} simplifies in this case and takes the form of the simplest real-field  $\phi^4$ sphaleron. 
Naturally, the boson profile solution for $\phi_1 $ is that  of the familiar $\phi^4$ kink,
\begin{equation}
	\phi_1^\text{sph}(z) = \frac{m}{2 \lambda} \tanh( \frac{m z}{2} ) \,, \quad
	{\rm while} \quad
	\phi_2^\text{sph}(z) \equiv 0 \,.
\label{MSTB_sphaleron}
\end{equation}
For $\abs{ \kappa } < 1 $, this solution is unstable with regards to decay into one of the two kinks \eqref{MSTB_kinks}; hence the name ``sphaleron''.
For a visual representation, see Figs.~\ref{fig:MSTB_kinks}, \ref{fig:MSTB_potential}, \ref{fig:ferm_eig}.
On the contrary, for $\abs{ \kappa } > 1 $, it is the only topological solution, as the two kinks \eqref{MSTB_kinks} disappear.
Below we will discuss this from the supersymmetric point of view.

\subsubsection{Semi-BPS state}

Let us have a closer look at the bosonic potential \eqref{MSTB_potential}. 
Setting $\phi_2 = 0$, we recover the familiar $\phi^4$ scalar potential.
Naively one would think that the kink in this model is a BPS state satisfying the first order equations and having (classically) the same central charge \eqref{zclas}.
However, in the case at hand it is also evident that, at least for  $\abs{ \kappa } < 1 $, this solution must possess larger energy than the kinks \eqref{MSTB_kinks}, since the kinks do saturate the central charge \eqref{zclas}, while the sphaleron is expected to be unstable.
So, what is going on?

The resolution of this puzzle lies in the non-holomorphic nature of the superpotential \eqref{MSTB_superpotential}.
Setting $\phi_2=0$ in this superpotential does not recover the familiar Wess-Zumino model with cubic superpotential; rather, it presents a one-parameter family of possible non-equivalent supersymmetrizations characterized by the parameter $\kappa$, see Appendix~\ref{sec:phi4_family} for a discussion.
All the models in this family have identical bosonic part of the action, but differ in the fermionic sector.

In particular, the trajectory \eqref{MSTB_sphaleron} passes through the point \eqref{W_singularity_1} in the field space which is a branch point of the square root in the superpotential.
There is no guarantee that the BPS mass formula would be valid for such a state, as the topological current \eqref{topcurr} or \eqref{ccdanoc} might not be conserved at the singularity.

And indeed, the first order equations satisfied by our sphaleron turn out to be
\begin{equation}
	\partial_z \phi_1 = \begin{cases}
		- \pdv{ \mathcal{W} }{ \phi_1 } \,, \quad &-\frac{m}{2 \lambda} < \phi_1 < -\kappa \frac{m}{2 \lambda} \,; \\[2mm]
		+ \pdv{ \mathcal{W} }{ \phi_1 } \,, \quad & -\kappa \frac{m}{2 \lambda} < \phi_1 < +\frac{m}{2 \lambda} \,. \\
	\end{cases}
\label{MSTB_first_order_eq_pm}
\end{equation}
As one can see, this is not quite a BPS equation, because of the discontinuity from the sign flip at a point.
Similar equations have been studied in the literature \cite{Alonso-Izquierdo:2023shi}; they have been called ``semi-BPS''.

The trajectory consists of two pieces, see Fig.~\ref{fig:MSTB_potential} for a visual representation.
This structure leads to the following formula for the sphaleron's energy (classically):
\begin{equation}
	M_\text{sph} 
	= \Big| \mathcal{W} (\text{Vac}_+ ) - \mathcal{W}(\text{irr}) \Big| + \Big| \mathcal{W}(\text{irr}) - \mathcal{W} ( \text{Vac}_- ) \Big|
	= \frac{ m^3 }{ 6 \lambda^2 } \,.
\label{MSTB_sphaleron_mass}
\end{equation}
Here, ``Vac$_\pm$'' are the two vacua \eqref{MSTB_vacua}, while ``irr'' denotes the irregular point \eqref{W_singularity_1} (in the present example, the superpotential happens to vanish at this point).
Of course, expression \eqref{MSTB_sphaleron_mass} does not depend on $\kappa$.
Note however that when the kinks \eqref{MSTB_kinks} do exist, i.e. in the range $-1 < \kappa < 1$, the kinks' energies \eqref{MSTB_kink_mass} are smaller than the sphaleron's energy \eqref{MSTB_sphaleron_mass}.

At the borderline case $\kappa = \pm 1$ the irregularity moves to the edge of the trajectory and merges with one of the vacua.
The two kinks \eqref{MSTB_kinks} and the sphaleron \eqref{MSTB_sphaleron} merge, and their rest energies \eqref{MSTB_kink_mass} and \eqref{MSTB_sphaleron_mass} become exactly the same.
This is the point where the ``semi-BPS'' sphaleron becomes a truly BPS state.

Beyond this threshold, $\abs{ \kappa } > 1 $, the would-be kink trajectories \eqref{MSTB_kinks} become imaginary, which means that they disappear.
The trajectory \eqref{MSTB_sphaleron} becomes the only stable topological solution (for uniformity, we will still call it a ``sphaleron'').
Naturally, the sign in \eqref{MSTB_first_order_eq_pm} becomes definite.
This topological soliton, now BPS, saturates the central charge \eqref{zclas},
\begin{equation}
	M_\text{sph} 
	= \abs{ \mathcal{W} (\text{Vac}_+ ) - \mathcal{W} ( \text{Vac}_- ) }
	= \frac{ m^3 }{ 6 \lambda^2 } \,.
\label{MSTB_sphaleron_mass_2}
\end{equation}
One should not be confused with the fact that seemingly the same difference of superpotentials yields different results in \eqref{MSTB_kink_mass} and \eqref{MSTB_sphaleron_mass_2}.
By inspecting the superpotential formulas \eqref{MSTB_vacua_Wval} one can see that they contain some absolute value signs which lead to different expressions depending on whether $\kappa$ is smaller or larger than $\pm 1$.

We stress that although the resulting energy formula $m^3/(6\lambda^2)$ in \eqref{MSTB_sphaleron_mass} and \eqref{MSTB_sphaleron_mass_2} appears to be the same, the unwrapping of this result in terms of the superpotential is very different.
Correspondingly, the physics of this solution is also very different, as will be discussed below.

\subsubsection{Bosonic modes}

Let us discuss the low-lying bosonic modes in the background \eqref{MSTB_sphaleron}.
First of all, for any value of $\kappa$ there is a translational zero mode,
\begin{equation}
	\delta\phi_a^{\omega=0} \sim \partial_z \phi_1^\text{sph}(z) \sim \delta_{1a} \frac{ 1 }{ \cosh[2]( \frac{mz}{2} ) } \,.
\label{sphaleron_zero_mode}
\end{equation}
For $|\kappa| > 1$ all the other modes have positive energy.

For $\abs{ \kappa } < 1 $, however, the sphaleron is unstable, and we should also see the corresponding negative bosonic mode.
This mode corresponds to the decay of the sphaleron into one of the kinks.
To find this mode, we need to consider the second-order equation of motion \eqref{bosonic_mode_equation_omega}.
Since the zero mode \eqref{sphaleron_zero_mode} has no nodes and is aligned with the $a=1$ flavor direction, the negative mode can be formed only in the ``transverse'' flavor direction $a=2$.
Thus, we need to study the equation
\begin{equation}
	- \partial_z^2 \delta\phi_2^\omega +  [\partial_2^2  U(\phi_1, \phi_2)] \delta\phi_2^\omega = \omega^2 \delta\phi_2^\omega \,.
\end{equation}
Note that $[\partial_1 \partial_2 U(\phi_1, \phi_2)] = 0$ on the sphaleron background, so that setting $\delta\phi_1 = 0$ is indeed consistent.
The problem at hand turns out to be a solvable P\"oschl-Teller problem, see also \cite{Alonso-Izquierdo:2018uuj}.
The only negative mode is given by 
\begin{equation}
	\delta\phi_a^\text{neg} = \delta_{2a} \frac{1}{ \cosh( \frac{mz}{2} ) } \,, \quad
	\omega^2 = - \frac{m^2}{4} (1 - \kappa^2) \,.
\label{sphaleron_negative_mode}
\end{equation}
Note a few things about this mode.
\begin{enumerate}
	\item This mode corresponds to the shift in the $a=2$ flavor. This is precisely as expected from a sphaleron ``sliding off'' downhill towards one of the kinks from the maximum in the scalar potential, see Fig.~\ref{fig:MSTB_potential_V}.
	\item Indeed, the frequency is negative, $\omega^2 < 0$ when $|\kappa| < 1$.
	\item Precisely at $\kappa = \pm 1$, when the two kinks \eqref{MSTB_kinks} and the sphaleron \eqref{MSTB_sphaleron} merge, this mode becomes a zero mode.
	As soon as $|\kappa| > 1$, we get $\omega^2 > 0$, and this mode is lifted to a positive energy.
\end{enumerate}

\subsubsection{Fermionic zero mode}
\label{sec:ferm_zero_mode_sphaleron_MSTB}

From the first-order equation for the sphaleron profile \eqref{MSTB_first_order_eq_pm} it is evident that for $|\kappa|<1$ the sphaleron is not quite supersymmetric.
However, at least at the classical level, this background breaks supersymmetry in a simple way.

\begin{figure}[t]
\centering
\begin{subfigure}[b]{0.9\textwidth}
	\centering
	\includegraphics[width=\textwidth]{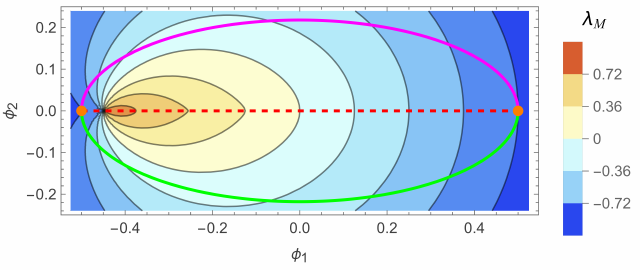}
	\subcaption{$\lambda_{\mathcal{M}}^{(1)}$}
	\label{fig:ferm_eig-1}
\end{subfigure}

\begin{subfigure}[b]{0.9\textwidth}
	\centering
	\includegraphics[width=\textwidth]{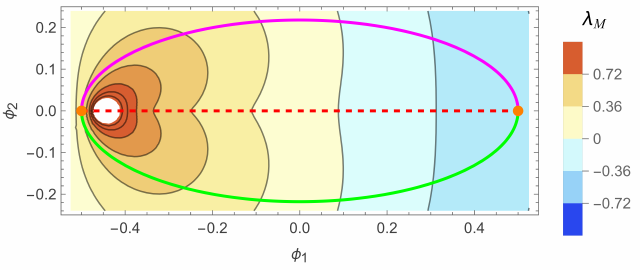}
	\subcaption{$\lambda_{\mathcal{M}}^{(2)}$}
	\label{fig:ferm_eig-2}
\end{subfigure}
\caption{\small
	Eigenvalues of the mass matrix \eqref{mm_1} as functions of bosonic fields $\phi_1,\phi_2$.
	For this plot, $\lambda = m = 1$ and $\kappa=0.9$.
	(\subref{fig:MSTB_potential_V}) First eigenvalue $\lambda_{\mathcal{M}}^{(1)}$.
	(\subref{fig:MSTB_potential_W1}) Second eigenvalue $\lambda_{\mathcal{M}}^{(2)}$.
	Labeling of the eigenvalues is the same as in Eq.~\eqref{mass_matrix_singularity}.
	Trajectories of the degenerate kinks \eqref{MSTB_kinks} are shown by solid lines, magenta and green respectively.
	Trajectory of the sphaleron \eqref{MSTB_sphaleron} is shown as a dashed red line.
	It is seen that the first eigenvalue $\lambda_{\mathcal{M}}^{(1)}$ is negative in both vacua, while the second eigenvalue $\lambda_{\mathcal{M}}^{(2)}$ changes sign along the soliton trajectories.
	Moreover, symmetry w.r.t. $\phi_2 \to -\phi_2$ is also evident.
}
\label{fig:ferm_eig}
\end{figure}

Let us start by discussing the zero modes at $|\kappa|<1$.
Applying SUSY transformations piecewise, we can still obtain the fermionic zero mode:
\begin{equation}
	\begin{pmatrix} \psi_1^a \\ \psi_2^a \end{pmatrix}
	=
	\delta_{1a} f(z) \epsilon(z) \,,
\label{tot_ferm_zero_mode_sphaleron}
\end{equation}
where the profile $f(z)$ is given by
\begin{equation}
	f(z) = \frac{ 1 }{ \cosh[2]( \frac{mz}{2} ) } \sim \partial_z \phi_1^\text{sph}(z) \,,
\end{equation}
while the spinor $\epsilon(z)$ is a step function:
\begin{equation}
	\epsilon(z) = \begin{cases} 
		{\scriptscriptstyle \begin{pmatrix} 1 \\ 0 \end{pmatrix} } \,, \quad z: \ \tanh( \frac{m z}{2} ) < -\kappa \,; \\[15pt]
		{\scriptscriptstyle \begin{pmatrix} 0 \\ 1 \end{pmatrix} } \,, \quad z: \ \tanh( \frac{m z}{2} ) > -\kappa \,. \\
	\end{cases}
\label{spinor_zero_mode_sphaleron}
\end{equation}
We have checked that this spinor indeed satisfies the corresponding Dirac equation \eqref{fermion_first_order_equations}.
The jump happens at the same point as the switch of the sign in the first order equation \eqref{MSTB_first_order_eq_pm}.
Recall that this is a direct consequence of the fact that the superpotential $\mathcal{W}$ has a square-root singularity, see Eq. \eqref{W_singularity_1}.
Near this point, the fermion mass matrix $\mathcal{M}_{ab} = \partial_a \partial_b \mathcal{W}$ is singular.
To the leading order, its eigenvectors and eigenvalues are given by\footnote{One should not be confused, as the two-component notation in \eqref{spinor_zero_mode_sphaleron} is for the spinor index $\alpha$, while the two-component notation in \eqref{mass_matrix_singularity} is for the flavor index $a$. }
\begin{subequations}
	\begin{equation}
		\mathcal{M}_{ab} v_b^{(i)} = \lambda_{\mathcal{M}}^{(i)} v_a^{(i)} \quad
		\text{(sum over $b$, no sum over $i$)} \,;
	\end{equation} 
	\begin{equation}
		\lambda_{\mathcal{M}}^{(1)} = \kappa m \cos\theta \,, \quad
		v_a^{(1)} = \big( \cos\theta \,, \ \sin\theta \big) \,,
		\label{mass_matrix_singularity_eq2}
	\end{equation}
	\begin{equation}
		\lambda_{\mathcal{M}}^{(2)} = \frac{(1 - \kappa^2) m^2}{4 \lambda} \frac{1}{\Upsilon} + \frac{1}{2} \kappa m \cos\theta \,, \quad
		v_a^{(2)} = \big( -\sin\theta \,, \ \cos\theta \big) \,,
	\end{equation}
	\begin{equation}
		\Upsilon \equiv \sqrt{ \left(\phi_1 + \kappa \frac{m}{2 \lambda}  \right)^2 + \phi_2^2 } \,, \quad
		\phi_1 = -\kappa \frac{m}{2 \lambda} + \Upsilon \cos\theta \,, \quad
		\phi_2 = \Upsilon \sin\theta \,.
	\end{equation}
\label{mass_matrix_singularity}
\end{subequations}
The dangerous eigenvalue is $\lambda_{\mathcal{M}}^{(2)}$: it diverges near this irregular point, see also Fig.~\ref{fig:ferm_eig}.
The fermionic mode corresponding to this eigenvalue is divergent (more on that below).
However, for the zero mode on the sphaleron background \eqref{MSTB_sphaleron} we have $\theta=0,\pi$, so that only the first eigenvalue $\lambda_{\mathcal{M}}^{(1)}$ is relevant.
The jump in this eigenvalue precisely accounts for the jump in the spinor \eqref{spinor_zero_mode_sphaleron}.

Now, when $\kappa \to \pm 1$, the would-be singularity is moved to one of the vacua, and the fermion mass matrix $\mathcal{M}_{ab}$ becomes smooth.
The spinorial part \eqref{spinor_zero_mode_sphaleron} of the zero mode \eqref{tot_ferm_zero_mode_sphaleron} becomes smooth; for example, at $\kappa \to 1$ this mode reduces to
\begin{equation}
	\begin{pmatrix} \psi_1^a \\ \psi_2^a \end{pmatrix}
	=
	\delta_{1a} \frac{ 1 }{ \cosh[2]( \frac{mz}{2} ) } \, \begin{pmatrix} 0 \\ 1 \end{pmatrix} \,.
\end{equation}

%%%%%%%%%%%%%%%%%%%%%%%%%%%%%%%%%%%%%%%%%%%%%%%%%%%%%%%%%%
%%%%%%%%%%%%%%%%%%%%%%%%%%%%%%%%%%%%%%%%%%%%%%%%%%%%%%%%%%
%%%%%%%%%%%%%%%%%%%%%%%%%%%%%%%%%%%%%%%%%%%%%%%%%%%%%%%%%%
%%%%%%%%%%%%%%%%%%%%%%%%%%%%%%%%%%%%%%%%%%%%%%%%%%%%%%%%%%
%\clearpage
\FloatBarrier

Precisely at $\kappa = \pm 1$ there is another normalizable mode for $a=2$,
\begin{equation}
	\begin{pmatrix} \psi_1^a \\ \psi_2^a \end{pmatrix}
	=
	\delta_{2a} \frac{ 1 }{ \cosh( \frac{mz}{2} ) } \, \begin{pmatrix} 0 \\ 1 \end{pmatrix} \,.
\label{sphaleron_negative_mode_psi}
\end{equation}
This fermionic zero mode is precisely the partner of the bosonic transverse mode \eqref{sphaleron_negative_mode} (recall that that bosonic mode is a negative mode and turns into a zero mode precisely at $\kappa = \pm 1$).
When $\kappa$ becomes large, $|\kappa| > 1$, the bosonic modes \eqref{sphaleron_negative_mode} and \eqref{sphaleron_negative_mode_psi} are lifted to positive-frequency modes.

\subsubsection{Would-be negative fermionic mode}
\label{sec:no_neg_ferm}

Now let us concentrate on the case $|\kappa| < 1$, when the sphaleron is unstable.
We would like to understand if there is a partner for the negative bosonic mode \eqref{sphaleron_negative_mode}.

If supersymmetry were (partially) unbroken in the sphaleron background, it would guarantee a fermionic partner for any non-zero bosonic mode.
However, if the sphaleron was to preserve a part of SUSY, it would be a BPS state and therefore saturate the lower bound on energy, which is never the case for an unstable state (because it decays into another state with lower energy).
Therefore, on general grounds we can conclude that an unstable state will break SUSY.

In the present case one of the manifestations of this breaking is the fact that there is no ``negative fermionic mode'' (whatever this term would mean), i.e. the negative bosonic mode \eqref{sphaleron_negative_mode} does not have a fermionic partner.
One way to see it is as follows.
We can try to go along the lines of the previous subsection and apply piecewise SUSY transformation \eqref{nonzero_fermionic_from_bosonic_modes} to the bosonic mode \eqref{sphaleron_negative_mode}.
Let us write them with explicit time dependence for Majorana fermions:
\begin{equation}
	\begin{pmatrix} \psi^a_1 \\ \psi^a_2 \end{pmatrix} = \begin{cases}
		\begin{pmatrix} - \cos(\omega t) \delta\phi_a \\[5pt] \frac{ \sin(\omega t) }{\omega} \mathcal{D}_{ab}^\dagger \delta\phi_b \end{pmatrix} \,, \quad z: \ \tanh( \frac{m z}{2} ) < -\kappa \,; \\[20pt]
		\begin{pmatrix} \frac{ \sin(\omega t)  }{\omega} \mathcal{D}_{ab} \delta\phi_b \\[5pt] \cos(\omega t) \delta\phi_a \end{pmatrix} \,, \quad z: \ \tanh( \frac{m z}{2} ) > -\kappa  \,.  \\
	\end{cases}
\label{nonzero_fermionic_from_bosonic_modes_piecewise}
\end{equation}
(plus one mode similar mode, cf. \eqref{nonzero_fermionic_from_bosonic_modes}).
One can see that for $\omega^2 > 0$ this mode would be oscillating, while for $\omega^2 < 0$ it indeed would become exponentially growing in time.
We have checked that the fermions \eqref{nonzero_fermionic_from_bosonic_modes_piecewise} do satisfy the Dirac equation \eqref{fermion_first_order_equations} provided that the bosons $\delta\phi_a$ satisfy the corresponding equation \eqref{bosonic_mode_equation_omega}.

Now, let us specifically consider the would-be negative mode which could be expected for $|\kappa|<1$.
The corresponding bosonic profile and $\omega$ are given by \eqref{sphaleron_negative_mode}.
Fermionic mode \eqref{nonzero_fermionic_from_bosonic_modes_piecewise} can be written down explicitly:
\begin{subequations}
	\begin{equation}
		\begin{pmatrix} \psi^a_1 \\ \psi^a_2 \end{pmatrix} = \delta_{2a} \times \begin{cases}
			\begin{pmatrix} - f(z,t) \\  g(z,t) \end{pmatrix} \,, \quad z: \ \tanh( \frac{m z}{2} ) < -\kappa  \,;  \\[15pt]
			\begin{pmatrix} - g(z,t) \\ f(z,t) \end{pmatrix} \,, \quad z: \ \tanh( \frac{m z}{2} ) > -\kappa  \,, \\
		\end{cases}
	\end{equation}
where
	\begin{equation}
		\begin{aligned}
			f(z,t) &= \cos(\omega t) \cdot \frac{1}{ \cosh( \frac{mz}{2} ) } \,, \\
			g(z,t) &= \frac{ \sin(\omega t)  }{\omega} \cdot \frac{1}{ \cosh( \frac{mz}{2} ) } \frac{( 1 - \kappa^2) m }{ 2 \left( \kappa   + \tanh( \frac{mz}{2} )  \right) } \,,
		\end{aligned}
	\end{equation}
	\begin{equation}
		\omega^2 = - \frac{m^2}{4} (1 - \kappa^2) \,.
	\end{equation}
\label{fermion_negative_mode_explicit}
\end{subequations}
Since $\omega^2 < 0$ for the case of interest $|\kappa|<1$, these profiles represent a solution exponentially growing in time.
However, precisely for $|\kappa|<1$, the function $g(z,t)$ is singular --- it has a pole at a finite value of $z$.
Therefore, fermionic mode \eqref{fermion_negative_mode_explicit} is not normalizable in $z$ (in the $L_2$ norm on the $z$-space $\mathbb{R}$).
This mode should be discarded.
The sphaleron does not have a ``negative fermionic mode''.

Note that the would-be fermionic solution became singular because of the singularity in the mass matrix $\mathcal{M}_{ab}$.
The bosonic mode avoids this singularity because the bosonic equation involves a non-trivial combination of the mass matrix elements, cf. eqs. \eqref{fermion_second_order_equations} and \eqref{DdagD}.
The singularity cancels in that combination.

From the explicit formula \eqref{fermion_negative_mode_explicit} one can also see that when the parameter $\kappa$ approaches $\pm 1$, the singularity in $g(z,t)$ runs away to $z \to \mp \infty$.
For $|\kappa| \geqslant 1$ there is no singularity anymore, $\omega^2 \geqslant 0$, and the fermionic mode becomes a normalizable fermionic partner of the bosonic mode \eqref{sphaleron_negative_mode}.
This perfectly agrees with the discussion at the beginning of this subsection: precisely at $|\kappa| \geqslant 1$ the sphaleron becomes stable and BPS, so all the non-zero modes are paired.

\subsection{Perturbative anomaly}
\label{sec:W_anom}

Before going into the non-perturbative effects, let us discuss implications of the perturbative anomaly reviewed in Sec.~\ref{sec:W_anomaly_general}.

\subsubsection{Correction to the superpotential}

Applying Eq.~\eqref{sevenan} to our superpotential from Eq.~\eqref{MSTB_superpotential}, we obtain
\begin{equation}
	\begin{aligned}
		{\mathcal W}_{\rm anom}
		&= \frac{1}{4 \pi} \partial_a \partial^a  {\cal W} \\
		&= \frac{1}{4 \pi} \left[ \frac{1}{\Upsilon} \left( \frac{m^2}{4\lambda} (1 + 2 \kappa^2) + \frac{3}{2} \kappa m \phi_1 \right) - 3 \lambda \Upsilon \right] \,.
	\end{aligned}
\label{W_anom_MSTB}
\end{equation}
Here, $\Upsilon$ is defined as in \eqref{MSTB_superpotential}.

Near the classical vacua \eqref{MSTB_vacua} we have $\phi_1 \sim \Upsilon \sim m/\lambda$ (assuming generic value of $\kappa \neq \pm 1$). 
The classical superpotential \eqref{MSTB_superpotential} and its second derivatives are of the order
\begin{equation}
	{\mathcal W}_{\rm class} \sim \frac{m^3}{\lambda^2} \,, \quad
	\partial_a \partial_b {\mathcal W}_{\rm class} \sim m \,.
\label{comparison_W_class}
\end{equation}
On the other hand, for the anomalous contribution \eqref{W_anom_MSTB} at the classical vacua we have
\begin{equation}
	{\mathcal W}_{\rm anom} \sim m \,, \quad
	\partial_a {\mathcal W}_{\rm anom} \sim \lambda \,, \quad
	\partial_a \partial_b {\mathcal W}_{\rm anom} \sim \frac{\lambda^2}{m} \,.
\label{comparison_W_anom}
\end{equation}
Comparing the classical result \eqref{comparison_W_class} with the quantum correction \eqref{comparison_W_anom} we see that, at least at the vicinity of the classical vacua, the quantum correction can be considered small, as long as we are at weak coupling $\lambda \ll m$ (and $\kappa \neq \pm 1$). 
In fact, the same statement holds also for the vicinity of the whole kink trajectories \eqref{ellipse_trajectory}.
Therefore, corrections to the kinks' masses \eqref{MSTB_kink_mass} are small.

Near the irregular point \eqref{W_singularity_1} where $\Upsilon \to 0$, this quantum correction to the superpotential becomes large.
This correction stays large even in the limit $\lambda \ll m$.
It disappears only for $\kappa = \pm 1$, when the degenerate kinks merge into a single kink.

In this paper we rely on the quasiclassical approximation, when the topological defects are treated classically.
The impact of the large quantum correction \eqref{comparison_W_anom} near the singular point will be studied in a subsequent publication in which all one-loop corrections will be systematically assembled.

\begin{figure}[t]
	\centering
	\includegraphics[width=0.7\textwidth]{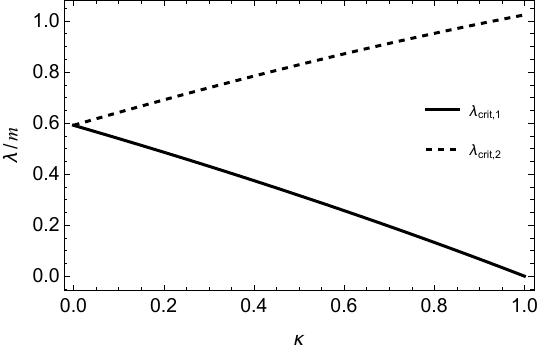}
	\caption{\small
		The first and the second critical values of the coupling, Eqs.~\eqref{lambda_crit_1} and \eqref{lambda_crit_2}.
		The quantum correction \eqref{W_anom_MSTB} does not spoil the theory as long as we stay below the solid line.
		Above this line, one of the vacua becomes lifted.
		Above the dashed line, both vacua are lifted.
	}
	\label{fig:MSTB_lambda_crit_2}
\end{figure}

\subsubsection{Scalar VEVs and the fate of the ground states}

Now let us consider some implications of the correction to the superpotential, Eq.~\eqref{W_anom_MSTB}.
We will focus on the case $|\kappa| \leqslant 1$ relevant for the study of degenerate kinks.
As in Eq.~\eqref{sevenan}, we introduce the full superpotential:
\begin{equation}
	\widetilde{\Wc} \equiv {\mathcal W}_{\rm class} + {\mathcal W}_{\rm anom} \,,
\label{W_quantum_corrected}
\end{equation}
where the classical piece ${\mathcal W}_{\rm class}$ is given by Eq.~\eqref{MSTB_superpotential}.
The scalar potential is determined by
\begin{equation}
	\widetilde{U} = \frac{1}{2} \left( \pdv{ \widetilde{\Wc} }{ \phi_a } \right)^2 \,.
\label{V_quantum_corrected}
\end{equation}

Quantum-corrected scalar VEVs in supersymmetric vacua are determined by the equations
\begin{equation}
	\pdv{ \widetilde{\Wc} }{ \phi_a } = 0 \,, \quad a=1,2
\label{vac_eq}
\end{equation}
One can show that in the quantum case, the vacua are still located at $\expval{\phi_2}=0$, while the VEV of $\phi_1$ is shifted from its classical value \eqref{MSTB_vacua}.
At small coupling $\lambda/m \ll 1$, the shifted values are given by
\begin{equation}
\begin{aligned}
	\text{Vac}_+ : \quad \expval{\phi_1} &= \frac{m}{2\lambda} - \frac{\lambda}{2 \pi m} \frac{2 + \kappa}{1 + \kappa} + O(\lambda^2/m^2) \,, \\
	\text{Vac}_- : \quad \expval{\phi_1} &= - \frac{m}{2\lambda} + \frac{\lambda}{2 \pi m} \frac{2 - \kappa}{1 - \kappa} + O(\lambda^2/m^2) \,.
\end{aligned}
\label{vac_shift_phi1}
\end{equation}
One can immediately see that the correction to $\text{Vac}_+$ in Eq.~\eqref{vac_shift_phi1} becomes dangerous if $\kappa \to -1$, while $\text{Vac}_-$ seems to be ill-behaved when $\kappa \to +1$.
This motivates us to take a closer look at these vacuum shifts.
It turns out, that both of these vacua are lifted at large enough value of the coupling constant $\lambda$.

In order to understand this better, let us simplify the vacuum equations \eqref{vac_eq}, assuming without loss of generality that $\kappa>0$.
The equation $\partial_2 \widetilde{\Wc} = 0$ is identically solved by $\phi_2=0$.
Then, if we assume $\kappa>0$ and focus on the vacuum at $\phi_1 < 0$, the equation $\partial_1 \widetilde{\Wc} = 0$ can be simplified to:
\begin{equation}
	f^2 - 1 + \frac{\lambda^2}{\pi m^2} \left( 3 +  \frac{1 - \kappa^2}{(f + \kappa)^2} \right) = 0 \,, \quad
	f \equiv \phi_1 \frac{2\lambda}{m}
\label{quartic_vac_eq_1}
\end{equation}
The last term here is the result of the quantum correction \eqref{W_anom_MSTB}.
Note that it is strictly positive, and for large enough $\lambda$ this equation might not have real solutions.

Eq.~\eqref{quartic_vac_eq_1} is equivalent to a quartic polynomial equation.
Behavior of its solutions can be studied e.g. by analysing the discriminant, see \cite{github-kink-inst} for details.
The result is the following.

When $\lambda$ is small enough, Eq.~\eqref{quartic_vac_eq_1} has two real solutions with $\phi_1 < - \kappa \frac{m}{2\lambda}$.
One of them corresponds to the shifted classical vacuum, Eq.~\eqref{vac_shift_phi1}.
The nature of the second solution (which formally also appears to present a supersymmetric vacuum) will be investigated in a subsequent work.
This second solution is situated closer to the singularity of ${\mathcal W}_{\rm anom}$, and for now we will ignore it.

\begin{figure}[t]
	\centering
	\begin{subfigure}[b]{0.48\textwidth}
		\centering
		\includegraphics[width=\textwidth]{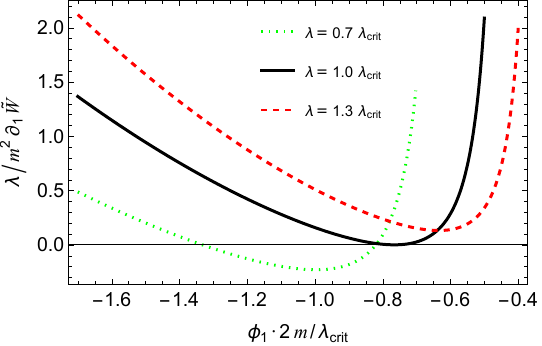}
		\subcaption{$\widetilde{\Wc}$}
		\label{fig:MSTB_near_crit_dWfull}
	\end{subfigure}
	~
	\begin{subfigure}[b]{0.48\textwidth}
		\centering
		\includegraphics[width=\textwidth]{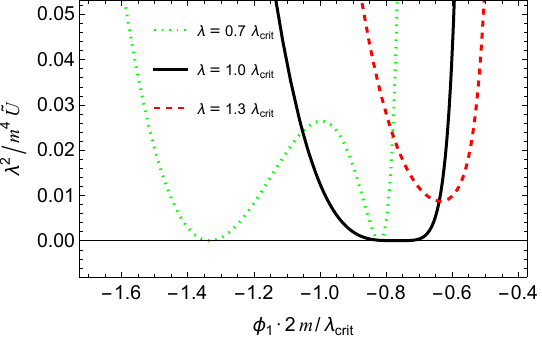}
		\subcaption{$\widetilde{U}$}
		\label{fig:MSTB_near_crit_Vfull}
	\end{subfigure}
	\caption{\small
		Quantum-corrected (super)potentials as functions of $\phi_1$ (slice $\phi_2=0$).
		(\subref{fig:phi4_W_kappa0}) shows the first derivative of the superpotential Eq.~\eqref{W_quantum_corrected} w.r.t. $\phi_1$, which is proportional to Eq.~\eqref{quartic_vac_eq_1}.
		(\subref{fig:phi4_W_kappa10}) shows the potential Eq.~\eqref{V_quantum_corrected}.
		For this plot, $\kappa = 0.4$.
		This figure shows the vicinity of one of the vacua, when the coupling is near the first critical value Eq.~\eqref{lambda_crit_1}.
		For the other vacuum and the coupling around Eq.~\eqref{lambda_crit_2}, the picture is similar.
	}
	\label{fig:MSTB_near_crit}
\end{figure}

Furthermore, when $\lambda$ hits a critical value (see Fig.~\ref{fig:MSTB_lambda_crit_2})
\begin{equation}
	\lambda_\text{crit,1} = \frac{m}{3} \sqrt{ \frac{\pi}{2} \left( 2 + \kappa^2 - \kappa \sqrt{ 3 (4 - \kappa^2) } \right)  } \,,
\label{lambda_crit_1}
\end{equation}
these two solutions merge.
For $\lambda > \lambda_\text{crit}$ these solutions disappear from the real line, and $\text{Vac}_-$ from Eq.~\eqref{vac_shift_phi1} becomes a local minimum of the scalar potential corresponding to a strictly positive vacuum energy, see Fig.~\ref{fig:MSTB_near_crit}.
We have checked (numerically) that, when the coupling constant approaches $\lambda_\text{crit,1}$, the Hessian matrix for the scalar potential $\widetilde{U}$ computed at $\text{Vac}_-$ has one positive and one zero eigenvalues.  

Repeating this study for the second vacuum at $\phi_1 > - \kappa \frac{m}{2\lambda}$, one can show that this one gets lifted when $\lambda$ reaches
\begin{equation}
	\lambda_\text{crit,2} = \frac{m}{3} \sqrt{ \frac{\pi}{2} \left( 2 + \kappa^2 + \kappa \sqrt{ 3 (4 - \kappa^2) } \right)  } \,.
	\label{lambda_crit_2}
\end{equation}
Note that $\lambda_\text{crit,1} < \lambda_\text{crit,2}$ when $\kappa > 0$, and vise versa.

We end this subsection with the following remark.
If we set $\kappa = 1$ the first critical value vanishes, $\lambda_\text{crit,1} = 0$.
In another limit $\kappa = - 1$, we get $\lambda_\text{crit,2} = 0$.
In either case, one of the vacua in the theory effectively becomes strongly coupled even at small $\lambda$.
Still, as long as $\kappa$ does not sit directly at $\pm 1$, there is a window of values of the coupling $\lambda$ such that the supersymmetric vacua stay intact.
This legitimizes the approximation studied in Sec.~\ref{sec:worldline_inst}.

\subsubsection{Perturbative correction to the kink mass}

By plugging the vacua \eqref{vac_shift_phi1} back into the superpotential $\widetilde{\Wc}$ from \eqref{W_quantum_corrected} and applying the BPS mass formula, we can derive the perturbative shift in the kink's mass as compared to the classical value \eqref{MSTB_kink_mass}.
This is justified as long as the coupling $\lambda$ stays below the first critical value \eqref{lambda_crit_1}.

\begin{figure}[t]
	\centering
	\includegraphics[width=0.7\textwidth]{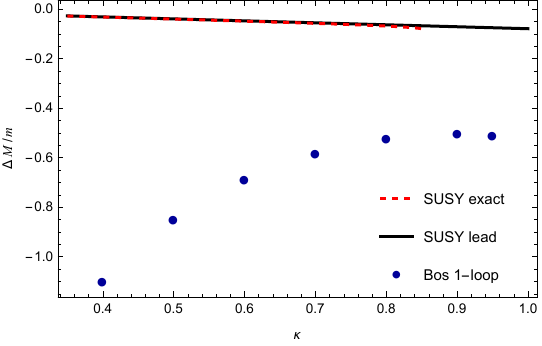}
	\caption{\small
		Quantum corrections to the kink mass (classical value Eq.~\eqref{MSTB_kink_mass} subtracted).
		Solid black like is the leading-order correction \eqref{MSTB_kink_mass_quant}, dashed red line is the exact result, blue dots represent the boson-only computation of Ref.~\cite{AlonsoIzquierdo:2002pit}.
	}
	\label{fig:MSTB_kink_mass_quantum_correction}
\end{figure}

The exact expression for the corrected mass is quite complicated, so here we present only the leading order correction:
\begin{equation}
	M_\text{kink} 
	= \frac{  m^3 }{ 12 \lambda^2 } |\kappa| (3 - \kappa^2)  - \frac{m |\kappa|}{ 4\pi } + O(\lambda^2/m^2) \,.
\label{MSTB_kink_mass_quant}
\end{equation}
This supersymmetric result can be compared to the contribution coming just from the bosonic sector; the latter was computed in Ref.~\cite{AlonsoIzquierdo:2002pit}, see Fig.~\ref{fig:MSTB_kink_mass_quantum_correction} here.

\vspace{10pt}

Corrections to the sphaleron mass \eqref{MSTB_sphaleron_mass} require careful analysis of all quantum effects in the vicinity of the singularity.
For now we will restrict ourselves to a comment on the quantum corrections coming just from the bosonic sector, computed in \cite{AlonsoIzquierdo:2002pit}.
For the sphaleron Eq.~\eqref{MSTB_sphaleron}, the classical mass  receives a correction from the bosonic fields
\begin{equation}
	[ \Delta M_\text{sph} ]_\text{bos} =  -\frac{ m}{\pi \sqrt{2}}\left[\left(3-\frac{\pi}{\sqrt{12}}\right)+\left(1-\sqrt{\kappa^2-1} \arcsin \frac{1}{\kappa}\right)\right] \,.
\label{sph_mass_correction_bosonic}
\end{equation}
For $|\kappa| < 1$, when the sphaleron is unstable, this correction has an imaginary part.

In the limit $\kappa \to 0$ both the real and imaginary parts of Eq.~\eqref{sph_mass_correction_bosonic} diverge as $\ln \kappa$.
This is directly related to the fact that at $\kappa = 0$ the O(2) of the potential is restored, and the discreteness of vacua is ruined.
Below in Sec.~\ref{sec:inst_bos_kappa-0} we will see that the same logarithmic divergence appears in the instanton action in this limit.

\subsection{Instanton (bosonic sector)}
\label{sec:instanton_MSTB}

Now it is time to discuss the instanton configuration that interpolates between the two degenerate kinks discussed above.
This study was initiated in \cite{Evslin:2025zcb}.
For now we will focus only on the bosonic sector; the fermionic sector will be discussed later.

\begin{figure}[t]
	\centering
	\includegraphics[width=0.4\textwidth]{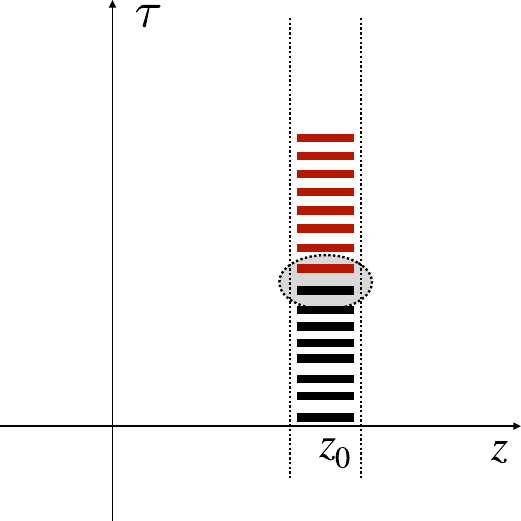}
	\caption{\small
		Instanton interpolating between kinks.
		The horizontal axis represents the spatial coordinate $z$, while the vertical axis is the Euclidean time $\tau$.
		Both kinks and the instanton are localized in space.
		At $\tau \to \pm \infty$, the instanton configuration tends to a kink; the two kinks are represented here by horizontal bars. The colors (red and black) represent the two degenerate types of kinks.
		At intermediate times, the instanton configuration interpolates between these two kinks.
		The instanton, localized both in space and in time, is represented here by a shaded circle.
	}
	\label{fig:misha_s_instanton_v02}
\end{figure}

The instanton is a solution of the Euclidean equations of motion
\begin{equation}
	\ddot{\phi}_a + \phi_a'' = \partial_a U(\phi_1,\phi_2) \,.
	\label{inst_bos_eq}
\end{equation}
The boundary conditions on the instanton solution $\phi_a^\text{inst}(z,\tau)$ are as follows:
\begin{equation}
	\begin{aligned}
		\lim_{\tau \to - \infty} \phi_a^\text{inst}(z,\tau) &= \phi_a^{\text{kink},-}(z) \,; \\
		\lim_{\tau \to + \infty} \phi_a^\text{inst}(z,\tau) &= \phi_a^{\text{kink},+}(z) \,; \\
		\lim_{z \to - \infty} \phi_a^\text{inst}(z,\tau)    &= \text{Vac}_- \,; \\
		\lim_{z \to + \infty} \phi_a^\text{inst}(z,\tau)    &= \text{Vac}_+ \,. \\
	\end{aligned}
	\label{instanton_bc_MSTB}
\end{equation}
Here, $\phi_a^{\text{kink},\pm}(z)$ are the two kink solutions in \eqref{MSTB_kinks}, while $\text{Vac}_\pm$ are the two vacua \eqref{MSTB_vacua}.
Thus, $\phi_a^\text{inst}(z,\tau)$ approaches the kinks asymptotically far in the Euclidean time $\tau \to \pm \infty$, see Fig.~\ref{fig:misha_s_instanton_v02} (cf. also Fig.~\ref{fig:instanton_winding}).

Physically, one way to think about an instanton is to consider it a classical trajectory (or a field profile) that interpolates between different \textit{ground states} of the theory.
At the same time, in the model at hand we wish to consider an instanton that interpolates between two \textit{kinks}.
To consolidate these notions, we can put our theory in boundary conditions with a spatial twist with respect to the $\mathbb{Z}_2$ transformation $S:\ \phi_1 \to - \phi_1$, see Eq.~\eqref{symmetries_mstb_S}.
Recall that this $S$ is a valid symmetry of the bosonic sector of the theory; the fermionic sector will be discussed below.
Such a twist puts the theory in the non-trivial topological sector; in particular, the ground states of this theory necessarily have non-trivial topology as well.
Since the two degenerate kinks are the global minima of the action with the given boundary conditions (they saturate the Bogomolny bound), we conclude that, in fact, these two kinks \textit{are} the ground states in this sector.

In other words, our instanton can be viewed as interpolating between two ground states in the twisted sector.
From the argument above we can also conclude that, when we twist with respect to $S$ at $z\to\pm\infty$, the $\mathbb{Z}_2$ symmetry $C:\ \phi_2 \to - \phi_2$ from Eq.~\eqref{symmetries_mstb_C} becomes spontaneously broken on the classical level\footnote{Translation symmetry is also broken in the kink background.}.
This can be seen from the fact that $C$ acts non-trivially on the degenerate kinks by interchanging them.
Quantum mechanically, in a non-supersymmetric theory, one could expect mixing of the two kinks via the instanton.
As a result, the $C$ symmetry is restored.
However, as we will see below, in the supersymmetric setup, the fermionic zero modes prevent this mixing, and $C$ stays spontaneously broken.

Unfortunately, supersymmetry is not going to be a lot of help in either computing the instanton action or finding the fermionic zero modes for the instanton.
It is immediately hinted by the fact that the $S$-transformation discussed above is not a symmetry of the fermionic sector.
Moreover, if we lift the model at hand to 2+1 dimensions, we can expect that the instanton becomes a static vortex%
\footnote{ In particular, at $\kappa \to 0$ this vortex approaches the global vortex configuration, see e.g. Sec.~3.1.1 of the textbook \cite{Shifman:2012zz}. This model is also relevant for Polyakov confinement, see e.g. Sec.~4.2 of \cite{Kovner:2000ss}. }%
.
If this vortex was a BPS state, one could expect to find at least two fermionic zero modes.
Therefore, such a vortex could be 1/2 BPS in the model with at least four supercharges, which is not the case at hand.
Therefore, we do not expect to find a Bogomolny representation for the instanton in 1+1D.

Still, we can study the instanton by other methods.
In particular, now we are going to provide a shortcut computation of the instanton action in the limit $\kappa \to \pm 1$, which was the focus of \cite{Evslin:2025zcb}.

\subsubsection{ \boldmath Bosonic worldline effective action at $\kappa \sim \pm 1$}
\label{sec:worldline_inst}

\paragraph{Derivation of the effective action.}

Near $\kappa \to \pm 1$ the trajectories of the two kinks \eqref{MSTB_kinks} pass very near each other in the field space.
Therefore, it makes sense that one can apply perturbation theory for studying the instanton interpolating between these kinks.
We will start with the case when the centers of these two kinks are aligned; translational invariance then allows us to put them at the origin $z=0$.
Later we will discuss what happens if the two kinks at $\tau \to \pm \infty$ are not aligned.

Just for this subsection let us switch to a more suitable notation by exchanging the $\kappa$ parameter to $\alpha$ (defined as in \cite{Evslin:2025zcb}),
\begin{equation}
	\kappa^2 = 1 - \alpha^2 \,.
\label{kappa_via_alpha}
\end{equation}
In terms of this parameter, the kink solution \eqref{MSTB_kinks} can be rewritten as
\begin{equation}
	\begin{aligned}
		\phi_1^\text{kink}(z) &= \frac{m}{2 \lambda} \tanh( \sqrt{1-\alpha^2} m z / 2 ) \,, \\
		\phi_2^\text{kink}(z) &= \pm \frac{m}{2 \lambda} \alpha \frac{1}{ \cosh( \sqrt{1-\alpha^2} m z / 2 ) } \,.
	\end{aligned}
\label{MSTB_kinks_alpha}
\end{equation}
The instanton configuration interpolates between these kinks, also passing through the sphaleron trajectory.
It minimizes the Euclidean action,
\begin{multline}
	S_E = \int dzd\tau \, \Bigg\{ \frac{1}{2} (\partial_\tau \phi_a)^2 + \frac{1}{2} (\partial_z \phi_a)^2 \\
		+ \frac{\lambda^2}{2} \left(\phi_1^2 + \phi_2^2 - \frac{m^2}{4 \lambda^2} \right)^2  + \frac{ ( 1 - \alpha^2 ) m^2 }{8} \phi_2^2
		\Bigg\} \,.
	\label{MSTB_action_eucl}
\end{multline}

Note that formally changing $\alpha \to - \alpha$ simply exchanges the two kinks in Eq.~\eqref{MSTB_kinks_alpha}, while setting $\alpha = 0$ we recover the sphaleron trajectory \eqref{MSTB_sphaleron}.
This motivates us to use the following ansatz for the instanton solution (inspired also by the results of \cite{Evslin:2025zcb}):
\begin{equation}
	\begin{aligned}
		\phi_1^\text{inst}(z,\tau) &= \frac{m}{2 \lambda} \tanh( \sqrt{1-\beta(\tau)^2} m z / 2 ) \,, \\
		\phi_2^\text{inst}(z,\tau) &= \pm \frac{m}{2 \lambda} \beta(\tau) \frac{1}{ \cosh( \sqrt{1-\beta(\tau)^2} m z / 2 ) } \,.
	\end{aligned}
\label{MSTB_inst_alphasmall}
\end{equation}
We will be interested in the limit of small $\alpha$.
The new quasimodulus --- the field $\beta(\tau)$ --- interpolates between $\pm \alpha$ at $\tau \to \pm \infty$.

With the ansatz \eqref{MSTB_inst_alphasmall}, it is straightforward to obtain the effective action for the $\beta(\tau)$.
To this end we plug the ansatz \eqref{MSTB_inst_alphasmall} into the Euclidean action \eqref{MSTB_action_eucl}, expand in the powers of $\beta$, and keep only a few relevant lowest-order terms.
Recall that we assume $|\beta(\tau)| < \alpha \ll 1$.
Integrating over $z$, we obtain%
\footnote{ See \cite{github-kink-inst} for details. } 
the effective quantum mechanical action for $\beta(\tau)$ (Euclidean notation):
\begin{equation}
	S_\beta^\text{eff} = \int d\tau \left\{ \frac{m}{\lambda^2} \frac{1}{2} \dot{\beta}^2 + V^\text{eff}(\beta) \right\} \,,
\label{beta_effact}
\end{equation}
\begin{equation}
	V^\text{eff}(\beta) \approx \frac{m^3}{6 \lambda^2 } - \frac{m^3 \alpha^2}{8 \lambda^2} \beta^2 + \frac{m^3}{ 16 \lambda^2 } \beta^4 \,.
\label{beta_effpot}
\end{equation}
The potential \eqref{beta_effpot} has minima at $\beta = \pm \alpha$.
The value of the potential at these points reproduces the kink energy \eqref{MSTB_kink_mass} with our level of accuracy,
\begin{equation}
	V^\text{eff}(\beta) \Big|_{\beta=\alpha} = \frac{m^3}{6 \lambda^2 } - \frac{m^3 \alpha^4}{16 \lambda^2} = \text{\eqref{MSTB_kink_mass}} + O(\alpha^6) \,.
\end{equation}
The point $\beta=0$ is an unstable stationary point corresponding to the sphaleron.
The value of the potential at that point $V(\beta=0)$ gives exactly the sphaleron energy \eqref{MSTB_sphaleron_mass}.
Of course, phenomenologically, the effective action \eqref{beta_effact} could have been written as soon as we knew the energies of the kinks and the sphaleron.

\paragraph{Instanton solution.}

The theory that we have built, Eq.~\eqref{beta_effact}, enjoys an instanton solution
\begin{equation}
	\beta(\tau) = \alpha \tanh( \frac{m \alpha}{2} \tau ) \,.
\label{beta_inst}
\end{equation}
By plugging this back into the initial ansatz \eqref{MSTB_inst_alphasmall}, we obtain the (approximate) instanton solution in the original MSTB model.

And here we come to an important point.
Since the potential $V(\beta)$ does not vanish at the minima, formally the corresponding instanton action is infinite.
In order to obtain a physically meaningful result for the instanton, one should discard the trivial piece $\int d\tau \, V^\text{eff}(\beta) \big|_{\beta=\alpha}$ corresponding to the vacuum energy.

This step is perfectly motivated from the 2d point of view as well.
Because a static kink has a constant energy, the kink's action (which involves integration over all of the time) is formally infinite.
Therefore, the action of a configuration involving an instanton interpolating between two kinks is going to be divergent as well.
What we are actually interested in is the \textit{relative} instanton action, that is, the action with the trivial piece (integral of kink's energy over time) subtracted away.

Performing the subtraction, we arrive at the result
\begin{equation}
	S_\text{inst} = \frac{ \sqrt{2} m^2 \alpha^3 }{ 3 \lambda^2 } \,.
\label{inst_action_MSTB}
\end{equation}
The instanton action \eqref{inst_action_MSTB} perfectly agrees\footnote{Note that our notation for $\lambda$ here differs from the mentioned paper: $\lambda_\text{\cite{Evslin:2025zcb}} = 2\lambda_\text{our}^2$, see also footnote \ref{ft:notation} on p. \pageref{ft:notation}.} with the result obtained in \eqref{MSTB_kinks_alpha} by solving the instanton equations of motion to a high order. % (see e.g. Eq. (5.20) of that paper).

\paragraph{\boldmath $S_\text{inst}$ via another shortcut.}

There is another easy way to compute the instanton action, once one knows the instanton profile at least to the lowest non-trivial order in $\alpha$.

Let us differentiate the Euclidean action \eqref{MSTB_action_eucl} with respect to the parameter $\alpha$.
One needs to differentiate only the explicit $\alpha$-dependence of the integrand, since all the other terms will be proportional to the equations of motions.
We have:
\begin{equation}
	\pdv{S}{\alpha} = - \frac{\alpha m^2}{4} \int dzd\tau \, (\phi_2)^2 \,.
\label{dSsAlpha_1}
\end{equation}
To compute the instanton action to the leading order in $\alpha$, we need to know only the leading-order approximation to the instanton solution.
Based on \eqref{MSTB_inst_alphasmall} and \eqref{beta_inst}, we can write:
\begin{equation}
	\begin{aligned}
		\phi_1^\text{inst}(z,\tau) &\approx \phi_1^\text{kink}(z) \,,
		\phi_2^\text{inst}(z,\tau) &\approx \phi_2^\text{kink}(z) \, \tanh( \alpha \frac{m \tau}{ 2 \sqrt{2} } )  \,,
	\end{aligned}
\label{phi_2_small_alpha}
\end{equation}
where $\phi_2^\text{kink}(z)$ is the static kink profile \eqref{MSTB_kinks_alpha}.
We will not need any higher-order terms.

The formula \eqref{dSsAlpha_1} by itself would give an integral divergent at large $\tau$, because it includes the kink's energy integrated over time.
To compute the relative instanton action, we need to subtract this contribution.
This amounts to computing the integral
\begin{equation}
	\pdv{S_\text{inst} }{\alpha} = - \frac{\alpha m^2}{4} \int dzd\tau \, \left( \phi_2^\text{kink}(z) \right)^2 \, \left( \tanh[2]( \alpha \frac{m \tau}{ 2 \sqrt{2} } ) -1 \right) \,.
\end{equation}
This gives,
for the (relative part of the) instanton action:
\begin{equation}
	\pdv{S_\text{inst} }{\alpha} =  \frac{ \sqrt{2} \alpha^2 m^2}{ \lambda^2 }  + O(\alpha^4) \,.
\label{dSsAlpha_2}
\end{equation}
This result can be easily integrated over $\alpha$. 
Demanding $S_\text{inst}(\alpha) \big|_{\alpha=0} = 0$ (because at this point the two kinks \eqref{MSTB_kinks_alpha} completely merge), we recover \eqref{inst_action_MSTB}.

\paragraph{Misaligned kinks.}

The effective action formalism also allows us to say what changes when the kinks are misaligned.
Namely, let us introduce a new coordinate $z_0(t)$ that parametrizes the center of the topological configuration.
Because of an overall translation invariance, $z_0$ is a zero mode, and the only additional contribution to the instanton's action comes from the kinetic term of $z_0(t)$, which is given by
\begin{equation}
	\Delta S_{z0} = \int d\tau \, \frac{m^3}{12 \lambda^2} (\dot{z}_0)^2 \,.
\end{equation}
Let us take the positions of the two kinks (between which the instanton is interpolating) to be $z_0^{-\infty}$ and $z_0^{+\infty}$.
The kinetic term for $z_0$ is minimized by the straight line trajectory.
Based on \eqref{beta_inst}, we take the instanton's span in the $t$-direction to be roughly
\begin{equation}
	- \frac{1}{m |\alpha|} < \tau < \frac{1}{m |\alpha|} \,.
\end{equation}
This gives
\begin{equation}
	\Delta S_{z0} \sim (z_0^{+\infty} - z_0^{-\infty})^2 \, \frac{m^4 |\alpha| }{24 \lambda^2} \,.
\label{misalignment_suppression}
\end{equation}
Thus we see that interpolation between misaligned kinks is exponentially suppressed.

\paragraph{Bosonic interference.}

Here, we have focused on the case when $\alpha$ is small, and basically we studied perturbation theory in $\alpha$.
However, if we allow $\alpha$ to be large, we can observe another interesting effect.

To see this effect, note that, if we constrain ourselves to the purely bosonic sector, the action in Eq.~\eqref{MSTB_action_eucl} is invariant under $\alpha \to - \alpha$.
At $\alpha = \pm 1$, the O(2) symmetry is restored, and the degenerate kinks \eqref{MSTB_kinks_alpha} are delocalized.
These facts allow us to identify the points $\alpha = \pm 1$; in fact, in our parametrization they correspond to the same value of $\kappa = 0$, see Eq.~\eqref{kappa_via_alpha}.

This means that the effective quantum mechanics \eqref{beta_effact} has a compact circle as a target space.
Since the masses of kinks and the sphaleron are formally smooth for all $\alpha \in [-1,1]$, we expect that the potential $V^\text{eff}(\beta)$ is also smooth on the entire target space.

Therefore, when $\alpha$ is not small, interference effects can become noticeable. 
In particular, there may be a certain value of the parameter $\alpha$, at which two different one-instanton amplitudes cancel because of the non-trivial relative phase.

\subsubsection{ \boldmath Instanton in the limit $\kappa \to 0$}
\label{sec:inst_bos_kappa-0}

Now we switch our notation back to $\kappa$ (recall that $\kappa = \sqrt{1 - \alpha^2}$).
In the limit $\kappa \to 0$, the last term in the potential \eqref{MSTB_potential} drops out locally, but still acts as a boundary condition.
The O(2) symmetry of the model becomes almost restored (it is broken by the boundary conditions only).

At the point $\kappa = 0$, the kinks disappear, but the instanton solution is formally there --- it becomes the global vortex.
Its action diverges logarithmically in this limit,
\begin{equation}
	S_\text{inst} = 2 \pi \frac{m^2}{4 \lambda^2} \int \frac{dr}{r} \,, \quad
	r = \sqrt{z^2 + \tau^2} \,.
\end{equation}
Here, $\tau$ is the Euclidean time.

Now let us keep $\kappa$ small but non-zero.
As seen from \eqref{MSTB_kinks}, the kink's width in the $z$-space blows up as
\begin{equation}
	z\text{-width} \sim \frac{1}{ \kappa m }  \sim \frac{1}{ \sqrt{1 - \alpha^2} m } \,.
\end{equation}
Because of the approximate O(2) symmetry, we expect that the instanton size in the $t$-direction is roughly the same.

This allows us to conclude that at small (but nonzero) $\kappa$, the instanton action is given by
\begin{equation}
	S_\text{inst} \sim \frac{ \pi m^2}{2 \lambda^2} \abs{ \ln \kappa }  \sim \frac{ \pi m^2}{4 \lambda^2} \abs{ \ln (1 - \alpha^2) } \,.
\label{s_inst_small-kappa}
\end{equation}
Note that the integral of the kink energy \eqref{MSTB_kink_mass} over the instanton's time span would give a negligible $O(1)$ contribution.
Therefore, the relative instanton action (with the kink contribution subtracted) is still given by \eqref{s_inst_small-kappa}.

One can make more precise statements about the instanton in this limit.
For definiteness, consider an instanton localized near the origin $\tau = z = 0$.
To find an instanton with topological charge 1, we can use the following ansatz:
\begin{equation}
	\phi_1 + i \phi_2 = \frac{m}{2\lambda} f( m r ) \, e^{i \theta} \,,
\end{equation}
where $r,\theta$ are the polar coordinates in the $z,\tau$ plane.
It is also convenient to introduce a dimensionless coordinate $\rho$ as
\begin{equation}
	r = \rho  / m \,.
\end{equation}
Plugging this ansatz into the instanton equations \eqref{inst_bos_eq} we arrive at
\begin{equation}
	f'' + \frac{1}{\rho} f' - \frac{1}{\rho^2} f - \frac{1}{2} (f^2 - 1) f = 0 \,.
\end{equation}
Such a system has been studied before, see e.g. 
\cite{Hindmarsh:1994re}. % Sec. 2.1. Global strings
The solution for $f$ exists and is smooth, with asymptotics
\begin{equation}
	f(\rho) \sim \rho - \frac{1}{16} \rho^2 \text{ at } \rho \to 0 \,, \quad
	f(\rho) \approx 1 - \frac{1}{\rho^2} \text{ at } \rho \to \infty \,.
\label{instanton_mstb_asymptotics}
\end{equation}
These asymptotics determine both the core and tail behavior of the instanton profile and can be used to match the near-core solution to the far-field logarithmic regime. 
In particular, they make explicit the origin of the logarithmic divergence in the action.

\subsection{Instanton zero modes}

Having discussed the bosonic part of the instanton solution, we turn now to the discussion of the fermionic zero modes.
The supersymmetry at hand does not allow us to reliably find all the fermionic zero modes.
One is faced with solving the Dirac equation in the instanton background.
However, even in this simple model, the Dirac equation is quite complicated.
In this section, we will give a few qualitative remarks.

\subsubsection{Overall translational modes of the kink-instanton system}
\label{sec:overall_ferm_mode}

We expect that the kink zero modes from Sec.~\ref{sec:ferm_kink_MSTB} smoothly continue through the instanton.
This is an overall fermionic mode of the configuration from Fig.~\ref{fig:misha_s_instanton_v02}.
When we switch on instantons, both in plain and topologically nontrivial backgrounds, we must perform the Wick rotation, i.e. pass 
to Euclidean time. Then one fermion zero mode on the kink we discussed  becomes a zero mode on the domain line in the plane $\{\tau,z\}$ ($\tau$ is the Euclidean time).
Its profile function depends on $z$ but is $t$-independent. The instanton deforms this line locally, near its center $z_0$ and $\tau_0$,
but the boundary behavior of the background field remains the same, which means that the profile is deformed  around $z_0,\tau_0$
but the mode survives. 

Thus, the configuration of two kinks plus instanton, Fig.~\ref{fig:misha_s_instanton_v02}, has a bosonic zero mode corresponding to the translation in $z$, and the corresponding fermionic%
\footnote{One might worry that the singularity in the mass matrix \eqref{mass_matrix_singularity} might pose an obstruction to having the renormalizable fermionic solutions of the Dirac equation. However, note that there is no singularity in the bosonic profiles; therefore, the singularity in the fermion mass matrix is of the form $\sim 1/r$ where $r = \sqrt{z^2 + \tau^2}$, and such a potential allows for well-behaved solutions \cite{Khalilov:1998jk}.} 
zero mode.
But these modes are not localized on the instanton. 
Those which are localized on the instanton cannot extend to large $t$.

\subsubsection{Modes localized on the instanton}

Now, the question is whether there are extra fermionic zero modes that actually are localized on the instanton both in $z$ and in $t$.
This question is very important, because mixing of kinks depends on it.
Indeed, in the purely bosonic case, the transition matrix element between the two degenerate kinks is non-zero due to the instanton.
In the present theory with a fermionic sector, if there are no fermionic zero modes localized on the instanton, then this transition amplitude is again non-zero.
This would mean that the kinks acquire corrections to their masses, and that the BPS saturation would be broken (at least for the antisymmetric combination of kinks).

Here we would like to argue that this mixing does not happen, as the transition amplitude vanishes due to the fact that there is a fermionic zero mode which is truly localized on the instanton.
This zero mode does not extend to the kinks, unlike the zero mode discussed in the previous subsection.

Unfortunately, index theorems cannot be directly applied in the present case, because one mass eigenvalue vanishes at a line in the field space, see footnote~\ref{ft:ferm_mass_zero} on p.~\pageref{ft:ferm_mass_zero}.
This does not allow us to rigorously prove our statement (unlike the $\mathcal{N}=2$ case treated in Sec.~\ref{sec:fermionic_modes_ads_inst}).
Still, below in Sec.~\ref{sec:mass_eig_signs_mstb} we present an argument, which, although conjectural, makes a plausible case.
But before doing that, we need to recall some facts about $\mathcal{N}=2$ 2d Wess-Zumino models with holomorphic superpotentials.

\subsubsection{ \boldmath Digression: fermionic zero modes in $\mathcal{N}=2$ models via $\mathcal{N}=1$ formalism}
\label{N=2_zero_modes_via_N=1}

To see that the situation outlined above is possible and even common, recall a familiar example of a $\mathcal{N}=(2,2)$ Wess-Zumino model with one complex superfield $\Phi$ and a holomorphic superpotential $\mathcal{W}_\text{WZ} (\Phi)$.
For example, a cubic superpotential leads to two vacua and a kink.
It is known that this kink has one complex (two Majorana) fermion zero mode.

Below we will follow Sec.~8 of \cite{Shifman:1998zy}.
The complex Wess-Zumino model can be written in terms of two real $\mathcal{N}=(1,1)$ superfields, $\Phi = \phi_1 + i \phi_2$.
In this notation, the superpotential becomes
\begin{equation}
	\mathcal{W} (\phi_1, \phi_2) = 2 \Im \mathcal{W}_\text{WZ} (\Phi) \,.
\label{W_from_Wwz}
\end{equation}
The fermion mass matrix is computed as above, $\mathcal{M}_{ab} = \partial_a \partial_b \mathcal{W}$.

Because the original superpotential $\mathcal{W}_\text{WZ}$ is holomorphic, the new superpotential $\mathcal{W}$ from \eqref{W_from_Wwz} is a harmonic function of $\phi_1$ and $\phi_2$.
Therefore, the fermion mass matrix $\mathcal{M}_{ab}$ is traceless, which leads us to the following statement: either $\mathcal{M}_{ab}$ has exactly one positive eigenvalue and one negative eigenvalue, or both of them vanish.

So, naively it would seem that the relative Morse index in the kink background vanishes, and fermionic zero modes are not guaranteed.
However, this is a bit more subtle.
In fact, the two eigenvalues are exchanged along the kink trajectory, in the sense that the corresponding eigenvectors switch places.
Thus, we in fact have two eigenvalues that change signs, thus yielding two real zero modes (one complex fermion).

To be more concrete, let us consider the most familiar example of a kink in a 2d Wess-Zumino model with the superpotential
\begin{equation}
	\mathcal{W}_\text{WZ} (\Phi) = \frac{m^2}{ 4\lambda} \Phi - \frac{\lambda}{3} \Phi^3 \,.
\end{equation}
This model has a single kink interpolating between vacua
\begin{equation}
	\Phi = \pm m/(2\lambda) \,,
\label{vacua_N=2_example}
\end{equation}
and this kink has one complex fermion zero mode.
In the $\mathcal{N}=1$ formalism, we can describe this model by the superpotential \eqref{W_from_Wwz},
\begin{equation}
	\mathcal{W} (\phi_1, \phi_2) = \frac{m^2}{ 4\lambda} \phi_2 + \frac{\lambda}{3} \phi_2^3 - \lambda \phi_1^2 \phi_2 \,.
\end{equation}
where we assumed $m,\lambda > 0$.
The mass matrix is just
\begin{equation}
	\mathcal{M} = 2 \lambda \begin{pmatrix}
		- \phi_2 & - \phi_1 \\
		- \phi_1 & \phi_2
	\end{pmatrix} \,.
\end{equation}
The eigenvalues and eigenvectors of this matrix are given by
\begin{equation}
	\begin{aligned}
		\lambda_{\mathcal{M}}^{(1)} &= -2 \lambda \Upsilon \,, \quad
		&v_a^{(1)} &= \big( \cos(\theta/2) \,, \ \sin(\theta/2) \big) \,;
		\\
		\lambda_{\mathcal{M}}^{(2)} &= + 2 \lambda \Upsilon \,, \quad
		&v_a^{(2)} &= \big( -\sin(\theta/2) \,, \ \cos(\theta/2) \big) \,,
	\end{aligned}
	\label{mass_matrix_N=2_example}
\end{equation}
where we used the parametrization
\begin{equation}
	\Upsilon \equiv \sqrt{ \phi_1^2 + \phi_2^2 } \,, \quad
	\phi_1 = \Upsilon \sin\theta \,, \quad
	\phi_2 = \Upsilon \cos\theta  \,.
\label{mass_matrix_N=2_example_parametrization}
\end{equation}
Note that as long as we stay away from the origin of the field space, the mass matrix always has one positive and one negative eigenvalue.
In particular, at the ground states \eqref{vacua_N=2_example}, the eigenvalues are $\pm m$ in each of the vacua; the corresponding eigenvectors are exchanged, as can be seen by comparing \eqref{mass_matrix_N=2_example} for $\theta = \pm \pi/2$.

We see that in this example, the relative Morse index between the two vacua \eqref{vacua_N=2_example} is zero.
In each of them, the mass matrix has two eigenvalues of opposite signs.
However, upon passing from one vacuum to the other (i.e. along the kink trajectory), the eigenvectors smoothly interpolate and eventually become exchanged.
This is the reason for the existence of the two fermionic zero modes.

\subsubsection{Tracing the mass eigenvalues in the MSTB instanton background}
\label{sec:mass_eig_signs_mstb}

\begin{figure}[t]
	\centering
	\includegraphics[width=0.6\textwidth]{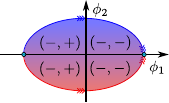}
	\caption{\small
		Signs of the eigenvalues of the fermion mass matrix $\mathcal{M}_{ab}$ on the field space spanned by $\phi_1$ and $\phi_2$.
		This figure can be viewed as a schematic depiction of Fig.~\ref{fig:ferm_eig} away from the singular point \eqref{W_singularity_1}.
	}
\label{fig:MSTB_signs}
\end{figure}

In the previous subsection we saw that, when an $\mathcal{N}=2$ supersymmetric system is viewed in the $\mathcal{N}=1$ formalism, fermionic zero modes may appear as a result of exchanged eigenvalues of opposite signs (or, alternatively, exchanged eigenvectors).
In fact, this is a necessity, because, as was argued above, due to holomorphicity of the superpotential, the relative Morse index always vanishes in a theory with a single complex superfield.

We would like to argue that a similar phenomenon occurs in the MSTB model under consideration, which is truly $\mathcal{N}=1$ and is not holomorphic.
Consider the signs of the mass eigenvalues on Fig.~\ref{fig:MSTB_signs}, which can be viewed as a simplified version of Fig.~\ref{fig:ferm_eig}.
In this figure, the shaded region is swept by the bosonic instanton solution interpolating from one kink to the other%
\footnote{Note that in this case the relative Morse index between the two vacua is not zero --- one eigenvalue changes sign, and this is the reason for the existence of one fermionic zero mode common to the kink in the instanton.
	This mode was discussed in Sec.~\ref{sec:overall_ferm_mode} above.
	Here we instead focus on zero modes localized on the instanton.}%
.

On the right side, roughly for $\phi_1 > - \kappa m / (2\lambda)$ (cf. Eq.~\eqref{W_singularity_1}), both mass eigenvalues are of the same sign.
When we pass from bottom $\phi_2 < 0$ to the top $\phi_2 > 0$, we do not have a fermionic mode supported in this region.

On the contrary, consider the left side of the field space, roughly at $\phi_1 < - \kappa m / (2\lambda)$.
In this case, the two mass eigenvalues have different signs.
As the example of Sec.~\ref{N=2_zero_modes_via_N=1} teaches us, we may have fermionic zero modes in this case.

Our argument is as follows.
The instanton interpolates between the kink in the lower half plane (at $\phi_2 < 0$) into the kink in the upper half plane (at $\phi_2 > 0$).
Under the transformation $\phi_2 \to - \phi_2$, the mass matrix transforms as
\begin{equation}
	\begin{aligned}
		\mathcal{M}_{11} &\to \mathcal{M}_{11} \,, \quad &\mathcal{M}_{12} &\to - \mathcal{M}_{12} \,, \\
		\mathcal{M}_{21} &\to - \mathcal{M}_{21} \,, \quad &\mathcal{M}_{22} &\to  \mathcal{M}_{22} \,.
	\end{aligned}
\end{equation}
Obviously, under this transformation the mass eigenvalues do not change, while the eigenvectors change sign of one of their components.
An example of this can be seen by examining Eq.~\eqref{mass_matrix_singularity} under $\theta \to - \theta$.

In the $\mathcal{N}=2$ case of Sec.~\ref{N=2_zero_modes_via_N=1}, this change of sign in the eigenvector components amounted to exchanging the eigenvectors, as we pass from one vacuum to the other.
In the present $\mathcal{N}=1$ case, this transformation does not simply switch the eigenvectors as we pass between the two kinks.
Still, we conjecture that it has the same effect: there are two Majorana zero modes localized on the kink.
In the field space, these zero modes are localized near a point on the l.h.s. of the shaded region in Fig.~\ref{fig:MSTB_signs}.

To illustrate, let us consider two limiting cases.

In the quasiclassical limit $\lambda / m \ll 1$ and at a generic $\kappa$ (not too close to 0 or $\pm 1$), the mass matrix has the structure (up to various $O(1)$ factors)
\begin{equation}
	\mathcal{M} \sim \begin{pmatrix}
		- \lambda \phi_1 & - \lambda \phi_2 \\
		- \lambda \phi_2 & m
	\end{pmatrix}	 \,.
\end{equation}
At the same level of accuracy, the mass eigenvalues and eigenvectors are given by:
\begin{equation}
	\begin{aligned}
		\lambda_{\mathcal{M}}^{(1)} &\sim - m \,, \quad &v_a^{(1)} &\sim (\lambda \phi_2  , -m ) \,; \\
		\lambda_{\mathcal{M}}^{(2)} &\sim - \lambda \phi_1  \,, \quad &v_a^{(1)} &\sim (m, \lambda \phi_2  ) \,. \\
	\end{aligned}
\end{equation}
One can see that roughly when $\lambda \phi_2 \sim \pm m$, the two eigenvectors are exchanged under the transformation $\phi_2 \to - \phi_2$.

In another limit, namely $\kappa \to 0$ and not-too-large $\lambda / m$ (below the critical lines on Fig.~\ref{fig:MSTB_lambda_crit_2}), the bosonic solution was discussed in Sec.~\ref{sec:inst_bos_kappa-0}.
In the notation of that section $\phi_1 + i \phi_2 = \frac{m}{2\lambda} f( m r ) \, e^{i \theta}$, the eigenvalues and eigenvectors of the mass matrix $\mathcal{M}_{ab}$ are given by (cf. \eqref{mass_matrix_singularity})
\begin{subequations}
	\begin{equation}
		\mathcal{M}_{ab} v_b^{(i)} = \lambda_{\mathcal{M}}^{(i)} v_a^{(i)} \quad
		\text{(sum over $b$, no sum over $i$)} \,;
	\end{equation} 
	\begin{equation}
		\lambda_{\mathcal{M}}^{(1)} = - m f + O(\kappa^2) \,, \quad
		v_a^{(1)} = \big( \cos\theta \,, \ \sin\theta \big) \,,
	\end{equation}
	\begin{equation}
		\lambda_{\mathcal{M}}^{(2)} = \frac{m}{ 2 f } \left( 1 - f^2 \right)  - \kappa \frac{m \cos\theta }{ 2 f^2 } + O(\kappa^2) \,, \quad
		v_a^{(2)} = \big( -\sin\theta \,, \ \cos\theta \big) \,.
	\end{equation}
\label{mass_matrix_kappa-0}
\end{subequations}
Here, $\theta$ is the polar angle in the field space $(\phi_1,\phi_2)$.
On our ansatz, $\theta$ coincides with the polar angle on the Euclidean coordinate plane $(z,\tau)$.
One can see that the eigenvectors are exchanged, if we compare two particular rays at $\theta = \pm 3\pi/4$.
We conjecture that the fermionic zero mode is localized near the line connecting the points of intersection of these two rays with the kink trajectories \eqref{ellipse_trajectory}.

\subsection{(No) mixing of kinks}

\subsubsection{Tunneling in the purely bosonic case}

Let us reiterate the key points of this Section.
The model under consideration, the MSTB model \eqref{MSTB_action}, displays two ground states and a pair of non-equivalent kinks \eqref{MSTB_kinks} interpolating between the two ground states.
Classically, these kinks are degenerate in energy.

In the purely bosonic case, on the quantum level, these kinks mix non-perturbatively due to the instanton discussed in Sec.~\ref{sec:instanton_MSTB}. The symmetric combination becomes the true kink, while the antisymmetric combination is lifted, acquiring an exponentially small mass correction.
As discussed in Sec.~\ref{sec:worldline_inst}, at least in the regime when $|\kappa| \sim 1$, the exponential splitting is given by
\begin{equation}
	\Delta E \sim \exp\left\{ \frac{ \sqrt{2} m^2 \alpha^3 }{ 3 \lambda^2 }  \right\} \,,
\end{equation}
where 
\begin{equation}
	\alpha = \sqrt{1 - \kappa^2} \ll 1 \,.
\end{equation}
When $\kappa$ is not close to $\pm 1$, one can have interference effects between different instantons.

If we denote $\ket{ \text{u-kink}}$ and $\ket{ \text{d-kink}}$ the quasiclassical wavefunctions corresponding to the two kinks in \eqref{MSTB_kinks}, then the ground state becomes an even combination $(\ket{ \text{u-kink}} + \ket{ \text{d-kink}})/\sqrt{2}$, while the exponentially split first excited states is the odd combination $(\ket{ \text{u-kink}} - \ket{ \text{d-kink}})/\sqrt{2}$.
The ``even'' and the ``odd'' here can be understood as the charges under the $\mathbb{Z}_2$ symmetry $C$ in Eq.~\eqref{symmetries_mstb_C}.
One can gauge this symmetry; in this case, the theory acquires a discrete $\theta$-angle and splits into two sectors, the even and the odd \cite{Dijkgraaf:1989pz}.
The ground state energies in the two sectors remain split --- one can say that the vacuum energy depends on $\theta$.
However, in the presence of fermions this $\theta$-dependence disappears, as we will argue below.

\subsubsection{Fermionic suppression}

After inclusion of fermions and the supersymmetrization discussed in Sec.~\ref{sec:MSTB_supersymmetrization}, not much changes on the classical level.
However, on the quantum level, the instanton acquires fermionic zero modes of its own.
We would like to argue that these modes suppress the tunneling amplitude, despite the fact that in the present theory even the fermion number is not well-defined (see Sec.~\ref{sec:F_loss} above).
As a result, the two kinks remain degenerate.

To see how this happens, let us recall the tunneling suppression in the supersymmetric quantum mechanics (see e.g. Sec.~2.3 of the textbook \cite{tongQM}).
The standard lore is as follows.
We consider quantum mechanics of one real scalar $x$ and two Grassmann fields $\psi$, $\psi^\dagger$ ($\sim$ two Majorana, or one complex fermion).
The Euclidean action is 
\begin{equation}
	S_E\left[x, \psi, \psi^{\dagger}\right] = \int d \tau\left[\frac{1}{2}\left(\frac{d x}{d \tau}\right)^2+\psi^{\dagger} \frac{d \psi}{d \tau}+\frac{1}{2} h^{\prime 2}-h^{\prime \prime} \psi^{\dagger} \psi\right] \,.
\end{equation}
If we take $h(x)$ to be a cubic polynomial, then the scalar potential $h^{\prime 2}/2$ is a quartic potential.
In the double-well case, the fermion mass $h^{\prime \prime}$ changes sign between two vacua, which is the reason for formation of a Grassmann zero mode of the instanton.
The path integral for the tunneling amplitude involves integration $\int d\eta$ over the corresponding collective mode (Grassmann number) $\eta$.
However, this integration is not saturated --- the integrand does not depend on $\eta$ (after all, it is a zero mode), and the whole integral vanishes.

Returning back to our model, we can make the following conjecture.
In the framework of the effective quantum described in Sec.~\ref{sec:worldline_inst}, after inclusion of fermions the action \eqref{beta_effact} should be supplemented by fermionic terms.
As was discussed above, our instanton breaks all the supersymmetries in the bulk, so we expect that the fermion coupling in the effective quantum mechanics is also non-supersymmetric.
Nevertheless, the fermionic zero mode of the instanton would manifest itself as a Grassmann collective mode, see also \cite{Jackiw:1975fn} for a related discussion.
The corresponding Grassmann integration leads to vanishing of the transition amplitude.

The above argument is qualitative. 
The case of minimal $\mathcal{N}=1$ supersymmetry is rather subtle. 
In the supersymmetrized MSTB model, a detailed analysis of the impact of fermion zero modes is warranted.
We plan to carry out this analysis in a subsequent publication.
When we pass to the extended $\mathcal{N}=2$ supersymmetry (see Sec.~\ref{sec:ADS_motivated_model}), the kink fermion number is well-defined and nullification of the instanton-related tunneling due to the fermions is obvious.

\section{ \boldmath ADS-motivated model with $\mathcal{N}=(2,2)$ supersymmetry }
\label{sec:ADS_motivated_model}

In this section we discuss degenerate kinks in the case of $\mathcal{N}=2$ supersymmetric models, focusing on a particular example motivated by 4D supersymmetric QCD (SQCD) with four supercharges.
Richer and more restrictive supersymmetry will allow us to make more rigorous statements, e.g., about the fermionic zero modes of the instanton interpolating between these kinks.

\subsection{Setup}

Our main example is a 2d Wess-Zumino model with a single chiral superfield $\Phi$ and the following superpotential:
\begin{equation}
	{\mathcal W} = \frac{a}{\Phi}+b\Phi \,.
\label{ADS_motivated_superpotential}
\end{equation}
Below, we will often use the two real components $\phi_a$, $a=1,2$, which we define as
\begin{equation}
	\Phi = \phi_1 + i \phi_2  \,.
\label{ADS_inst_ReIm_parametrizarion}
\end{equation}
The parameters in the superpotential \eqref{ADS_motivated_superpotential}, $a$ and $b$, are in principle two arbitrary complex numbers.
For simplicity, we choose them to be real and positive; a generalization to complex $a,b$ is straightforward.
The (bosonic part of the) action reads:
\begin{equation}
	S = \int dz \, dt \,  \left\{ |\partial_\mu \Phi|^2 - V  \right\} \,, \quad
    V = \abs{ \pdv{\mathcal W}{\Phi} }^2 = \abs{ b - \frac{a}{ \Phi^2 } }^2
\label{action_integral_1}
\end{equation}
This model has two vacua, as determined by the critical points of $\mathcal{W}$:
\begin{equation}
	{\mathcal W}' = 0 
	\Rightarrow
	\Phi_\pm = \pm \sqrt{ \frac{a}{b} } \,.
\label{vacua}
\end{equation}
Here, the prime represents a derivative w.r.t. $\Phi$.
Perturbative mass near each of the vacua \eqref{vacua} as calculated form the scalar potential $V$ in Eq.~\eqref{action_integral_1}:
\begin{equation}
    (m_\text{pt})^2 = \frac{4 b^3}{a} \,.
\label{m_pert}
\end{equation}

As we will see below, this model enjoys two degenerate kinks interpolating between these vacua.
Much like in the previous section, we want to consider a possibility of having an instanton interpolating between these kinks.

The superpotential \eqref{ADS_motivated_superpotential} is motivated by $\mathcal{N}=1$ SQCD in 4D, with gauge group $SU(2)$ and one quark flavor. 
There, $\Phi$ plays the role of the meson superfield. 
In that context, $b=m$ is the quark mass, and $a=\Lambda^5$ in terms of the strong coupling scale. 
Generalization to the gauge group $SU(N)$ yields the superpotential
\begin{equation}
	\mathcal{W}_N = \tilde{a} \Phi +  \frac{1}{N-1} \frac{ \tilde{b} }{ \Phi^{N-1} } \,.
\end{equation}
In this paper we will focus on the simplest case $N=2$.
Note that the kinetic term in Eq.~\eqref{action_integral_1} is not exactly what it would have been in the context of SQCD, but this is not important for the purposes of this paper.

Let us briefly discuss the global symmetries of this model, which closely resemble the case discussed in the previous section, see Eq.~\eqref{symmetries_mstb}.
There are two internal $\mathbb{Z}_2$ symmetries which we will call $S$ and $C$, acting as
\begin{subequations}
	\begin{equation}
		S: \quad \phi_1 \to - \phi_1 \,, \text{ or } \Phi \to - \bar{\Phi} \,;
		\label{symmetries_ads_S}
	\end{equation}
	\begin{equation}
		C: \quad \phi_2 \to - \phi_2 \,, \text{ or } \Phi \to \bar{\Phi} \,.
		\label{symmetries_ads_C}
	\end{equation}
	\label{symmetries_ads}
\end{subequations}
The $S$ symmetry is again spontaneously broken, as seen from the ground states in Eq.~\eqref{vacua}.
The difference with respect to the MSTB model of Sec.~\ref{sec:mstb} is that here the $S$ symmetry is respected by the fermionic sector.

\subsection{Two degenerate kinks}

Let us start by describing the bosonic profiles for the two degenerate kinks.
We will derive them from the Bogomolny representation for the energy of static field configurations.

The static energy functional, as follows from \eqref{action_integral_1}, is given by
\begin{equation}
	{\mathcal E} = \int dz \left\{  \abs{ \partial_z\Phi }^2 + \abs{ b - \frac{a}{ \Phi^2 } }^2 \right\} \,.
\label{energy_static_second}
\end{equation}
We can rewrite it in the Bogomolny form:
\begin{equation}
	{\mathcal E} = \int dz \left| \partial_z\Phi \pm \frac{\partial \bar{\mathcal W}}{\partial\bar\Phi}	\right|^2 + \mathcal{Z}  \,,
\label{energy_static_BPS}
\end{equation}
where $\mathcal{Z}$ is a fixed number --- the value of the central charge. 
For a BPS kink saturating the lower bound on the energy, we obtain the first order profile equation
\begin{equation}
	\partial_z\Phi = \pm \frac{\partial \bar{\mathcal W}}{\partial\bar\Phi}
	= \pm \left( b - \frac{a}{ \bar{\Phi}^2 } \right) \,.
\label{1st_order_eq}
\end{equation}
For a kink satisfying the boundary conditions
\begin{equation}
	\lim_{z \to - \infty} \Phi(z) = \Phi_- \,, \quad
	\lim_{z \to + \infty} \Phi(z) = \Phi_+ \,,
\label{bc}
\end{equation}
(cf. \eqref{vacua}), the central charge is given by
\begin{equation}
	\mathcal{Z} = 2 [ {\mathcal W}(\Phi_+) - {\mathcal W}(\Phi_-) ] = 8 \sqrt{ab} \,.
	\label{kink_central_charge}
\end{equation}

Equation \eqref{1st_order_eq} can be solved straightforwardly.
It turns out that there are two types of solutions with the boundary conditions \eqref{bc}, two kinks and a sphaleron; the latter is discussed below.

In order to derive the kink solutions, following \cite{Kovner:1997ca,Ritz:2004mp}, let us first make a substitution $\Phi = \Phi_+ e^{i \theta}$ in the BPS equation \eqref{1st_order_eq}:
\begin{equation}
	\partial_z \theta = - 2 \frac{b^{3/2}}{a^{1/2}} \sin\theta \,.
\end{equation}
This equation is solved by
\begin{equation}
	\theta = \pm 2\; \mathrm{arctan} ( e^{- m_\text{pt} \, (z - z_0) } ) \,,
\end{equation}
which yields the two kink solutions
\begin{subequations}
	\begin{equation}
		\Phi_\text{u-kink}(z) = \Phi_+ \cdot \frac{ e^{m_\text{pt} \, z} + i }{ e^{m_\text{pt} \, z} - i } \,,
		\label{kink_solutions_u}
	\end{equation}
	\begin{equation}
		\Phi_\text{d-kink}(z) = \Phi_+ \cdot \frac{ e^{m_\text{pt} \, z} - i }{ e^{m_\text{pt} \, z} + i } \,.
		\label{kink_solutions_d}
	\end{equation}
\label{kink_solutions}
\end{subequations}
Here, $m_\text{pt}$ is naturally the mass of perturbative excitations near each of the vacua, see Eq.~\eqref{m_pert}.
In terms of the real component fields \eqref{ADS_inst_ReIm_parametrizarion}, the two kinks become
\begin{equation}
    \phi_1(z) = \Phi_+ \tanh( m_\text{pt} \, z ) \,, \quad
    \phi_2(z) = \pm \frac{ \Phi_+ }{ \cosh( m_\text{pt} \, z ) } \,,
\label{kink_solutions_ads_real}
\end{equation}
cf. Eq.~\eqref{MSTB_kinks}. The plus/minus sign in Eq.~\eqref{kink_solutions_ads_real} corresponds to the u/d-kink respectively.

\begin{figure}[t]
	\centering
	\includegraphics[width=0.5\textwidth]{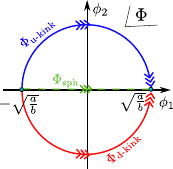}
	\caption{\small
		Complex plane of $\Phi$ (cf. Fig.~\ref{fig:MSTB_kinks} from the previous Section).
		Vacua \eqref{vacua} are shown by two cyan dots.
		Kinks \eqref{kink_solutions} are shown by the blue (upper) and red (lower) solid lines respectively.
		The would-be sphaleron \eqref{sphaleron_implicit} is shown by green dashed line going through the center.
	}
\label{fig:ADS_kinks}
\end{figure}

Both kinks in Eq.~\eqref{kink_solutions} interpolate from $\Phi_-$ at $z \to - \infty$ to $\Phi_+$ at $z \to + \infty$, see Fig.~\ref{fig:ADS_kinks}.
The first kink $\Phi=\Phi_\text{u-kink}(z)$ goes through the upper half-plane on the field space; the phase of $\Phi$ (i.e. $\theta$) interpolates from $+\pi$ to $0$ clockwise.
The second kink $\Phi=\Phi_\text{d-kink}(z)$ goes through the lower half-plane on the field space; the phase of $\Phi$ interpolates from $-\pi$ to $0$ going counter-clockwise.

Each kink is a 1/2 BPS state and has one complex fermion zero mode (two Majorana modes).
Note also that $\Phi_\text{u-kink}(z) = \overline{\Phi_\text{d-kink}(z)}$, i.e. the two kinks are exchanged by the $\mathbb{Z}_2$ transformation $C$ from Eq.~\eqref{symmetries_mstb_C}.

\subsection{Sphaleron}

We want to consider a sphaleron solution.
Due to the $\mathbb{Z}_2$ symmetry $C:\ \Phi \to \bar{\Phi}$ the sphaleron trajectory in the $\Phi$ complex plane must follow along the real axis.
One could expect that, for such a trajectory, the first term in \eqref{energy_static_BPS} might not be zero, but in fact it vanishes, and naively the sphaleron would appear to be BPS as well.
Below we are going to show this.

\subsubsection{First order equations for sphaleron profile}

Since the sphaleron typically represents an unstable maximum of the action, one can expect that the energy of such a configuration would be higher than the kink's mass.
Naively, this means that the sphaleron trajectory cannot possibly satisfy the first order BPS equations, otherwise it would entail \eqref{energy_static_BPS}.
However, it turns out that the first order equation \eqref{1st_order_eq} has another, real-valued, solution.

To find this solution, we demand that $\Im \Phi = 0$, so that $\Phi = \bar{\Phi}$. 
After this substitution, one can easily integrate \eqref{1st_order_eq} and find a solution in an implicit form:
\begin{equation}
	z - z_0 = \frac{1}{b} \left[ \Phi_+ \, \text{atanh} \frac{\Phi_\text{sph}(z)}{\Phi_+} - \Phi_\text{sph}(z)  \right] \,,
\label{sphaleron_implicit}
\end{equation}
where the vacuum value $\Phi_+$ is defined in \eqref{vacua}, and $z_0$ is an arbitrary constant (the sphaleron center).
For a visual representation, see Fig.~\ref{fig:ADS_kinks}.
Note that, contrary to a kink, the sphaleron \eqref{sphaleron_implicit} satisfies Eq.~\eqref{1st_order_eq} with the choice of sign as ``$-$''.

The function $\Phi_\text{sph}(z)$ implicitly defined by Eq.~\eqref{sphaleron_implicit} is continuous everywhere, and differentiable everywhere except at $z = z_0$.
In the vicinity of this point, we have
\begin{equation}
	\Phi_\text{sph}(z) \approx (3 a (z-z_0))^{1/3} \,, \quad
	\Phi_\text{sph}'(z) \sim \frac{1}{ (z-z_0)^{2/3} } \,.
\label{ADS_sphaleron_singularity}
\end{equation}
Thus $\Phi_\text{sph}(z)$ satisfies the first order BPS equation almost everywhere.
Plugging this solution into the energy functional \eqref{energy_static_second} we see that it diverges, thus signifying that the sphaleron energy is infinite and this configuration is unphysical.

Thus we have a trajectory that satisfies the first order BPS equation but at the same time is very far from saturating the Bogomolny bound \eqref{energy_static_BPS}.

\subsubsection{Sphaleron energy and BPS bound}

To sort this out, let us revisit the formula for the BPS central charge and particle mass \eqref{kink_central_charge}.
The standard lore is to write down the Bogomolny representation which automatically yields
\begin{equation}
	\mathcal{Z} = 2 \int dx \, \partial_z \Phi \pdv{ \mathcal{W} }{\Phi} = 2 [ {\mathcal W}(\Phi_+) - {\mathcal W}(\Phi_-) ] \,.
\label{central_charge_2}
\end{equation}
This formula  gives $\mathcal{Z}_\text{kink} = 8 \sqrt{ab}$ for the kink.
Naively, the result for the sphaleron appears to be the same.

Now, recall that the superpotential in the model under consideration, Eq.~\eqref{ADS_motivated_superpotential}, is singular at $\Phi = 0$.
The differential $d {\mathcal W}$ is not well-defined at that point.
There is no guarantee that the BPS mass formula would be valid for such a state, as the conservation of the topological current \eqref{topcurr} or \eqref{ccdanoc} can be violated there.

In fact, let us analyze the integral \eqref{central_charge_2} more closely.
If we regularize it by cutting out an interval $[-\epsilon, \epsilon]$ around zero, we obtain 
\begin{equation}
	\mathcal{Z}_\text{kink} + \frac{4a}{\epsilon} \,.
\end{equation}
We could have deformed the initial integration contour by going a small half-circle around the origin, and this would have cancelled the divergent piece.
This is what happens for the kink, whose trajectory goes around the origin.
However, the sphaleron must pass right through the origin, and this makes the sphaleron's energy not well-defined (formally it's infinite).
This situation is similar to the ones studied in \cite{Manton:2023afn,Alonso-Izquierdo:2023shi}.

Let us conclude with a remark on the translational zero mode for this sphaleron.
In principle, it should be given by
\begin{equation}
	\delta\Phi(z)_\text{transl} \sim \partial_z \Phi_\text{sph}(z) \,.
\end{equation}
However, because of the singularity \eqref{ADS_sphaleron_singularity} we find that $\phi(z)_\text{transl} \sim (z-z_0)^{-2/3}$, and the translational mode is not normalizable.
This is another manifestation of the fact that the field configuration $\Phi_\text{sph}(z)$ has infinite rest energy.

\subsection{Instanton action and bosonic profiles}

Much like in Sec.~\ref{sec:instanton_MSTB} above, we want to discuss the instanton in the present model.
The instanton configuration $\Phi_\text{inst}$ interpolates between the two degenerate kinks, and we impose the following boundary conditions on the Euclidean plane $(z,\tau)$:
\begin{equation}
	\begin{aligned}
		\lim_{\tau \to - \infty} \Phi_\text{inst}(z,\tau) &= \Phi_\text{u-kink}(z) \,; \\
		\lim_{\tau \to + \infty} \Phi_\text{inst}(z,\tau) &= \Phi_\text{d-kink}(z) \,; \\
		\lim_{z \to - \infty} \Phi_\text{inst}(z,\tau) &= \Phi_-  \,; \\
		\lim_{z \to + \infty} \Phi_\text{inst}(z,\tau) &= \Phi_+ \,. \\
	\end{aligned}
	\label{instanton_BC}
\end{equation}
Here, $\Phi_\text{u/d-kink}$ are the two kink solutions from \eqref{kink_solutions}.
In a similar way as discussed in Sec.~\ref{sec:instanton_MSTB}, we can think of this instanton as interpolating between two ground states of the theory with spatial boundary conditions twisted by the $S$ transformation Eq.~\eqref{symmetries_ads_S}

We will now begin by writing down the Bogomolny representation for the instanton and computing its action.
Eq.~\eqref{inst_act} presenting the final answer explicitly demonstrates that the instanton action is finite.
The question of the fermion zero modes will be addressed later.

\begin{figure}[t]
	\centering
	\includegraphics[width=0.5\textwidth]{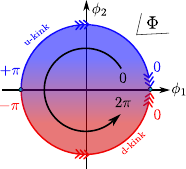}
	\caption{\small
		Instanton interpolating between kinks, drawn on the field space (complex plane of $\Phi$).
		The u-kink \eqref{kink_solutions_u} (the upper arch shown in blue) winds in phase from $+\pi$ to 0.
		The d-kink \eqref{kink_solutions_u} (the lower arch shown in red) winds in phase from $-\pi$ to 0.
		The instanton configuration in our convention winds by $2\pi$, as shown by the solid black arrow in the middle.
	}
	\label{fig:instanton_winding}
\end{figure}

\subsubsection{Topological charge}

Let us start by writing the Bogomolny representation for the instanton action.
This is possible because the instanton of the 1+1D model under consideration can be naturally lifted to a vortex in 2+1D or a two-wall junction in 3+1D.
In the derivation below we will closely follow \cite{Gorsky:1999hk}.

We will frequently use the following parametrization (polar coordinates in the field space):
\begin{equation}
	\Phi = \Upsilon \cdot e^{i \theta} \,.
\label{theta_parametrization}
\end{equation}

From the Euclidean time $\tau$ and the spatial $z$ coordinates, we form complex coordinates $\zeta$, $\bar{\zeta}$ defined as
\begin{equation}
	\zeta = z + i\tau \,, \quad
	\partial_\zeta = \frac{1}{2} ( \partial_z - i \partial_\tau )  \,.
\label{complex_coordinates_spacetime}
\end{equation}
We choose the orientation on the Euclidean plane $(z,\tau)$ as set by the canonical volume form $dz \wedge d\tau$.

The Bogomolny representation is obtained by rewriting the action as
\begin{equation}
	\begin{aligned}
		S 
		&= \int dz \, d\tau \,  \left\{ |\partial_\mu \Phi|^2 + \abs{ \pdv{ \bar{\mathcal{W}} }{ \bar{\Phi} } }^2 \right\} \\
		&= \int dz \, d\tau \,  %\left\{ 
		\abs{ 2 \partial_\zeta \Phi - \pdv{ \bar{\mathcal{W}} }{ \bar{\Phi} } }^2  
        %+ S_\text{kink}^{(1)} + S_\text{kink}^{(2)} 
        + \int d\tau M_\text{kink}
		+ S_\text{inst}
        \,.
		%\right\}
	\end{aligned}
\label{ADS_inst_bogomolny_rep}
\end{equation}
%
% Here, the terms $S_\text{kink}^{(1,2)}$ come from $2 \int dz \, d\tau \, [\partial_\zeta \mathcal{W} + c.c.]$; they correspond to the central charges of the static kinks between which the instanton interpolates\footnote{Formally these terms are divergent, as we need to integrate the kink central charges over all of time. However, with given boundary conditions at $z \to \pm \infty$, a kink represents a ground state, so we subtract that contribution away, as in Sec.~\ref{sec:mstb}.}.
%
Here, $M_\text{kink}$ is given by the absolute value of the central charge \eqref{kink_central_charge}.
The term $\int d\tau M_\text{kink}$ in \eqref{ADS_inst_bogomolny_rep} comes from $2 \int dz \, d\tau \, [\partial_\zeta \mathcal{W} + c.c.]$.
Formally, this term is divergent, as we need to integrate the kink mass over all of time. However, with given boundary conditions at $z \to \pm \infty$, a kink represents a ground state, so we subtract that contribution away, as in Sec.~\ref{sec:mstb}.

The term $S_\text{inst}$ in \eqref{ADS_inst_bogomolny_rep} gives the instanton action; it is given by
\begin{equation}
	\begin{aligned}
		S_\text{inst}
		&= 2 \int dz \, d\tau \,  \left[ \partial_\zeta \bar{\Phi} \partial_{\bar{\zeta}} \Phi - \partial_\zeta \Phi  \partial_{\bar{\zeta}} \bar{\Phi} \right] \\
		&= 2 \int dz \, d\tau \, ( \dot{\phi}_1 \phi_2' - \dot{\phi}_2 \phi_1' ) \\
		&= \frac{1}{2} \int a^\mu \, dx_\mu \,,
	\end{aligned}
\label{vortex_central_charge}
\end{equation}
where
\begin{equation}
	\begin{aligned}
		a^\mu 
		&= i \left( \bar{\Phi} \partial^\mu \Phi - \Phi \partial^\mu \bar{\Phi} \right) \\
		&= 2 \left( \phi_2 \partial^\mu \phi_1 - \phi_1 \partial^\mu \phi_2 \right) \\
		&= -2 \Upsilon^2 \partial^\mu \theta \,,
	\end{aligned}
\label{vortex_central_charge_current}
\end{equation}
see \eqref{theta_parametrization}\footnote{For various signs in \eqref{vortex_central_charge} and \eqref{vortex_central_charge_current}, it is important that our choice of orientation is set by $dz \wedge dt$. From that one can easily check all signs in \eqref{ADS_inst_bogomolny_rep}--\eqref{vortex_central_charge_current} on a simple example $\Phi(z,\tau) = z + i\tau$. }.
Thus, if we consider an instanton with the boundary conditions specified in \eqref{instanton_BC}, we obtain
\begin{equation}
	\int a^\mu \, dx_\mu = 2 |\Phi_+|^2 \cdot (\pi - (-\pi)) \,,
\end{equation}
see also Fig.~\ref{fig:instanton_winding}.
This gives the instanton action
\begin{equation}
	S_\text{inst} = 2 \pi |\Phi_+|^2 = 2 \pi \frac{a}{b} \,.
\label{inst_act}
\end{equation}

The integrand in \eqref{vortex_central_charge} is nothing but the $(1/2, 1/2)$ central charge corresponding to the axial current. 
From the last line of \eqref{vortex_central_charge_current} we see that this central charge counts the winding of the phase of the complex field $\Phi$.
This winding number (with a minus sign) is exactly the topological charge of the instanton at hand.

Recall that a solitary kink has a translational zero mode corresponding to the location of the kink's center $z_0$.
In principle, the two kinks (between which our instanton is interpolating) could be positioned at different points, $z_0^{(1)}$ and $z_0^{(2)}$ respectively.
Because of the overall translation symmetry, the instanton action can depend on these centers' coordinates at most through their difference $z_0^{(1)} - z_0^{(2)}$.
This goes for any translation-invariant theory.

Naively, in our case it seems like the instanton action \eqref{inst_act} does not depend on $z_0^{(1)} - z_0^{(2)}$ at all, since the integral of the central charge \eqref{vortex_central_charge} can be taken on a large contour with the size much larger than $| z_0^{(1)} - z_0^{(2)} |$.
However, we expect that when the two kinks are misaligned and $z_0^{(1)} - z_0^{(2)} \neq 0$, the instanton just does not saturate the central charge --- its action is larger as compared to Eq.~\eqref{inst_act}.
This is similar to domain walls in higher dimensions: formally, the boundary conditions and the BPS central charge are unchanged if the wall is somehow bent or inflated, but such a wall simply would not saturate the BPS bound.

We also make the following remark.
Typically, in 2+1D a solitary vortex has energy that diverges in the IR.
It can be regularized e.g. by introducing an anti-vortex somewhere in the system (then the IR cutoff is determined by the vortex-antivortex distance), or by embedding in an Abelian Higgs model (where the IR cutoff is determined by the photon's mass).

In the present case, if we lift the present model to 2+1D, the instanton under consideration becomes a vortex-like junction of two domain walls.
The energy of the vortex is simply proportional to the action \eqref{inst_act} and is finite.
However, it is finite only because we compute the energy counting off the kink (domain wall) background.

\subsubsection{BPS profile equations and the singularity}

From the Bogomolny representation \eqref{ADS_inst_bogomolny_rep} we can read off the first order equations:
\begin{equation}
	2 \partial_\zeta \Phi = \pdv{ \bar{\mathcal{W}} }{ \bar{\Phi} } = b - \frac{a}{ \bar{\Phi}^2 } \,.
\end{equation}
In terms of parametrization \eqref{ADS_inst_ReIm_parametrizarion} we have
\begin{equation}
	\begin{cases}
		\phi_1' + \dot{\phi}_2 &= b - a \frac{ \phi_1^2 - \phi_2^2 }{ (\phi_1^2 + \phi_2^2)^2 } \,; \\
		\phi_2' - \dot{\phi}_1 &= - 2 a \frac{ \phi_1 \chi_2 }{ (\phi_1^2 + \phi_2^2)^2 }  \,. \\
	\end{cases}
\end{equation}
In terms of parametrization \eqref{theta_parametrization} we have
\begin{equation}
	\begin{cases}
		\Upsilon' + \Upsilon \, \dot{\theta} &= \left( b - \frac{a}{\Upsilon^2} \right) \cos\theta \,; \\
		\Upsilon \, \theta' - \dot{\Upsilon} &= - \left(b + \frac{a}{\Upsilon^2} \right) \sin\theta \,. \\
	\end{cases}
\label{inst_eq_ads_fieldpolar_coordcartesian}
\end{equation}
Note that the degenerate kinks from Eq.~\eqref{kink_solutions} also satisfy these equations (assume $\tau$-independent fields and compare Eq.~\eqref{inst_eq_ads_fieldpolar_coordcartesian} with the kink equation \eqref{1st_order_eq}).

Let us introduce polar coordinates on the Euclidean plane $(z,\tau)$ according to
\begin{equation}
	z = r \cos\alpha \,, \quad
	\tau = r \sin\alpha \,.
\label{polar_coordinates}
\end{equation}
In terms of these polar coordinates, the holomorphic derivative \eqref{complex_coordinates_spacetime} can be rewritten as
\begin{equation}
	2 \partial_\zeta = e^{- i \alpha } \left( \partial_r - \frac{i}{r} \partial_\alpha \right) \,.
\end{equation}
Then,
\begin{equation}
	\begin{cases}
		\Upsilon_r + \frac{1}{r} \Upsilon \theta_\alpha &= b \cos(\alpha - \theta) - \frac{a}{\Upsilon^2} \cos(\alpha + \theta) \,; \\
		\Upsilon \theta_r - \frac{1}{r} \Upsilon_\alpha &= b \sin(\alpha - \theta) - \frac{a}{\Upsilon^2} \sin(\alpha + \theta) \,. \\
	\end{cases}
\label{ADS_instanton_Ftheta_polar}
\end{equation}
Previously we saw that the sphaleron had a singularity, which led to a divergence of the sphaleron's energy.
One might worry, what is the fate of the instanton, since the instanton configuration inevitably passes through the point $\Phi=0$ in the field space; let's say that this happens at the origin of the $(z,\tau)$ plane.
From the form \eqref{ADS_instanton_Ftheta_polar} it is easy to see that in the vicinity of this point we have, to the leading order,
\begin{equation}
	\Upsilon \approx \left( \frac{3}{2} a r \right)^{1/3} \,, \quad
	\theta \approx - \alpha \,, \quad
	\Phi_\text{inst} \approx \left( \frac{3}{2} a \right)^{1/3} \frac{ z - i \tau }{ (z^2 + \tau^2)^{1/3} } \,,
\end{equation}
where $\Upsilon, \theta$ is the parametrization \eqref{theta_parametrization}, and $r,\alpha$ are the polar coordinates in the $(z,\tau)$ Euclidean plane \eqref{polar_coordinates}.
Plugging this into the action in the first line of Eq.~\eqref{ADS_inst_bogomolny_rep},
we find that the would-be divergence of the action integral
\begin{equation}
	S \sim \int \frac{ r \, dr }{ r^{4/3} }
\label{2d_convergence}
\end{equation}
actually becomes convergent due to the fact that the integration measure is now two-dimensional.

\subsection{Fermionic zero modes}

Now let us discuss fermionic zero modes of the topological solutions discussed in this Section, starting with the two degenerate kinks \eqref{kink_solutions}.

\subsubsection{Overall modes}

Since in our setup we have $\mathcal{N} = (2,2)$ supersymmetry (four supercharges in 2d) and the kinks are 1/2 BPS, each one of them conserves two supercharges.
It is easy to see that they actually conserve \textit{the same} pair of supercharges.
The two remaining supercharges generate two real fermionic zero modes (one complex mode) for each kink.
The fermion mass $\mathcal{W}''$ winds in the kink's background, which is the reason for zero-mode formation.

However, the zero modes just discussed are not localized on the instanton --- they are common to all Euclidean ``wall'' configurations, which is the same as in Fig.~\ref{fig:misha_s_instanton_v02}.
This was discussed in Sec.~\eqref{sec:overall_ferm_mode} in the context of the MSTB model.

\subsubsection{Modes localized on the instanton}
\label{sec:fermionic_modes_ads_inst}

We would like to argue that there are two real fermionic zero modes (one complex) localized on the instanton under consideration.

\vspace{10pt}

This can be easily seen from a higher-dimensional perspective.
Indeed, for a moment let us consider the same theory embedded into 2+1D spacetime.
Each of the kinks then becomes a static domain wall.
The instanton interpolating between these kinks becomes a two-wall junction vortex.
Evidently, such a vortex is also 1/2 BPS.
It breaks two supercharges that generate two fermionic zero modes localized on the vortex.
Compactifying this theory on a small circle, we return to the 2d setup.

An interesting phenomenon is seen when we inspect these fermionic zero modes more closely.
First, from the point of view of SUSY multiplets, these two modes are paired with the two translational modes corresponding to moving the vortex (i) along the wall in the $\tau$-direction, and (ii) in the perpendicular $z$-direction together with the wall.
Naively these two fermionic modes are Majorana; however, the wall's translational mode should be complex, because, initially, the kink in 1+1D had two Majorana zero modes partnered with the translational bosonic mode.
The resolution of this puzzle seems to be the supersymmetry enhancement analogous to the phenomenon described in \cite{Ritz:2004mp}.
We expect that both fermionic zero modes are complexified.
The one that is paired with $\tau$-translations becomes a complex mode localized on the junction vortex.

One concern could be that the fermion zero modes are not normalizable because of the singularity of the superpotential at $\Phi=0$.
However, this is not the case, as can be seen as follows.
From the SUSY transformations, fermion zero modes are given by $\psi \sim \slashed{\partial} \Phi_\text{inst}$, where $\Phi_\text{inst}$ is the bosonic solution.
The norm of the zero mode then becomes
\begin{equation}
	\int dzd\tau \, \bar{\psi} \psi \sim \int dzd\tau \, | \partial \Phi_\text{inst} |^2 \,,
\end{equation}
which converges by the same argument that led to \eqref{2d_convergence}.

\vspace{10pt}

Note that this argument could not be applied in the $\mathcal{N}=(1,1)$ model from Sec.~\ref{sec:mstb}.
In 2+1D, a vortex with winding number one always has two zero modes.
Thus, a vortex can be 1/2 BPS only in a theory with four (or more) supercharges, see also the discussion in Sec.~II.A of \cite{Ritz:2004mp} or in Sec.~2.2.2 of \cite{Shifman:2009zz}.

The fermionic zero modes can also be counted with the help of the index theorems for the Dirac operator \cite{Callias:1977kg,Weinberg:1981eu,Jackiw:1981ee}.
These index theorems demand, among other things, that the fermion mass (or the coefficient in front of the $\psi\psi$ or $\bar{\psi}\psi$ term) is non-vanishing as long as we are sufficiently far from the vortex location.
This is satisfied in the present case, because the fermion mass winds in the complex plane on the kink's background \eqref{kink_solutions}, staying away from zero.
However, in the model from Sec.~\ref{sec:mstb}, the fermion mass actually vanishes at one point on the kink, arbitrarily far from the instanton-vortex, see footnote~\ref{ft:ferm_mass_zero} on p.~\pageref{ft:ferm_mass_zero}.

\subsection{Suppression of the tunneling amplitude}

Thus, we are led to the conclusion that the instanton under consideration has one localized fermionic zero mode.
Unlike for the theory considered in Sec.~\ref{sec:mstb}, in the present case the fermion number is well-defined.
The supercharge is characterized by fermion number 1, and the same goes for each fermionic zero mode of the instanton.

Therefore, if the instanton starts from a purely bosonic kink at $\tau \to -\infty$, then at $\tau \to +\infty$ it must arrive at a two-fermion kink (both fermionic modes filled).
But this state is orthogonal to the bosonic kink.
We conclude that the tunneling amplitude between the two kinks is suppressed in the $\mathcal{N}=(2,2)$ supersymmetric setup.

At the same time, if one considers a purely bosonic theory, this transition amplitude is non-vanishing, because the instanton action \eqref{inst_act} is finite.

\section{Conclusions}
\label{sec:concl}

In this paper we have studied models with degenerate kinks, and investigated whether these kinks can mix via an instanton.
Our main result is that, while these kinks do mix in a bosonic theory, adding supersymmetry brings in fermionic zero modes that are localized on the instanton, thus completely suppressing the tunneling amplitude between the two kinks.

Supersymmetrization of the MSTB model turned out to be non-trivial, with a superpotential that has a root-like irregularity in the field space.
This model has an unstable semi-BPS topological configuration (a sphaleron), which can become stable and fully BPS in a certain region of the parameter space.
We also studied the fate of the would-be negative fermionic mode of the sphaleron: it is non-normalizable when the sphaleron is unstable, but becomes a normalizable positive-energy mode when the sphaleron is stable.
We believe that such a mismatch of bosonic-fermionic modes might be typical for semi-BPS configurations.

The $\mathcal{N}=2$ model studied in this paper also has degenerate kinks and a singularity of the superpotential.
The sphaleron configuration in this case, while naively being BPS, turns out to have infinite energy; thus, it does not really exist.
The instanton configuration, however, still exists and has a finite action.

Beyond their intrinsic interest, these findings provide insight into the interplay between solitons, instantons, and unstable classical solutions in models with multiple scalar fields. 
The methods we developed can be applied more broadly to other two–dimensional theories with nontrivial vacuum structure, and suggest further lines of research on the correspondence between kink–like objects and semiclassical tunneling phenomena, as well as on models with non-analytic superpotentials.

\section*{Acknowledgments}

We thank Shi Chen and Aleksey Cherman for helpful discussions.
This work is supported in part by U.S. Department of Energy Grant No. de-sc0011842.

\appendix

\section{ \boldmath One-parameter family of $\mathcal{N}=1$ super\-sym\-me\-tri\-za\-tions of the $\varphi^4$ model}
\label{sec:phi4_family}

In this Appendix we report on one curious byproduct of this study.
Let us start by recalling the action for a two-dimensional $\varphi^4$ model with a single real scalar field:
\begin{equation}
	S_\text{bos} = \int dzdt \, \left\{ \frac{1}{2} (\partial_\mu \varphi)^2 - U(\varphi) \right\} \,, \quad
	U(\varphi) = \frac{\lambda^2}{2} \left(\varphi^2 - \frac{m^2}{4 \lambda^2} \right)^2  \,.
\label{action_phi4_bos}
\end{equation}
Normalization of the couplings here is not important (it is chosen for convenience).

The standard way of supersymmetrizing this model is to write down the superpotential, which we will call $\mathcal{W}_\infty$ (notation will become clear in a moment):
\begin{equation}
	\mathcal{W}_\infty (\varphi) = \frac{m^2}{ 4\lambda} \varphi - \frac{\lambda}{3} \varphi^3 \,.
\label{superpotential_phi4_standard}
\end{equation}
One can check that $ U = \frac{1}{2} | \mathcal{W}_\infty ' |^2 $, where prime denotes the derivative with respect to $\varphi$.
As reviewed in Sec.~\ref{sec:N=11_generalities}, the fermionic part of the action reads
\begin{equation}
	S_\text{ferm} = \int dzdt \, \left\{ \bar\psi \,i\gamma^\mu \partial_\mu \psi 
	- \frac{1}{2} \,	\mathcal{W}_\infty '' \, \bar\psi \psi \right\}  \,.
\label{action_phi4_ferm}
\end{equation}
Note that the full action is symmetric under $\varphi \to - \varphi$, while the superpotential \eqref{superpotential_phi4_standard} transforms as $\mathcal{W}_\infty \to - \mathcal{W}_\infty$.

\begin{figure}[t]
	\centering
	\begin{subfigure}[b]{0.48\textwidth}
		\centering
		\includegraphics[width=\textwidth]{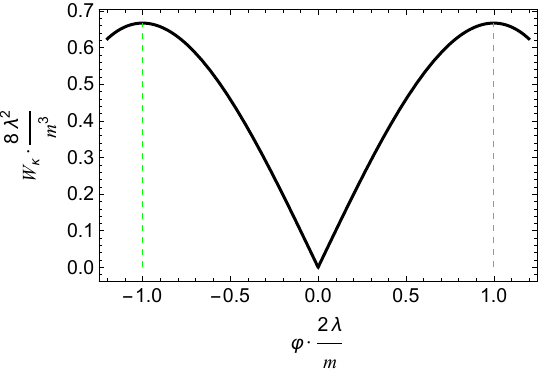}
		\subcaption{$\kappa = 0$}
		\label{fig:phi4_W_kappa0}
	\end{subfigure}
	~
	\begin{subfigure}[b]{0.48\textwidth}
		\centering
		\includegraphics[width=\textwidth]{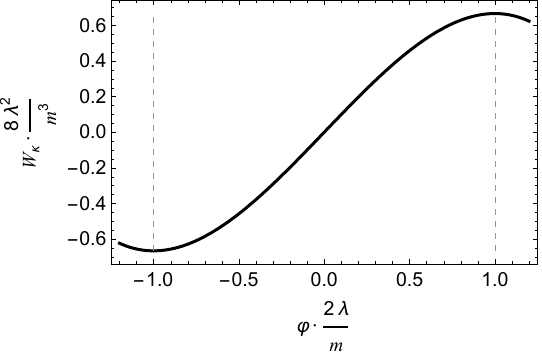}
		\subcaption{$\kappa \ll m/\lambda$}
		\label{fig:phi4_W_kappa10}
	\end{subfigure}
	\caption{\small
		Examples of superpotentials from the family \eqref{superpotential_phi4_family_1}-\eqref{superpotential_phi4_family_2}.
		Thick black line is the superpotential, dashed green lines denote the field values corresponding to the two ground states.
		(\subref{fig:phi4_W_kappa0}) shows $\kappa = 0$ with a non-analyticity of the superpotential.
		(\subref{fig:phi4_W_kappa10}) shows large $\kappa$ (with an unimportant constant in \eqref{superpotential_phi4_family_2} subtracted away), when the superpotential is smooth at finite field values.
	}
	\label{fig:phi4_W}
\end{figure}

Turns out, there is a one-parameter family of superpotentials that yield the same bosonic potential as in Eq.~\eqref{action_phi4_bos}.
Consider the following expression:
\begin{equation}
	\mathcal{W}_\kappa (\varphi) = \frac{m^2}{ 4\lambda} \Upsilon - \frac{\lambda}{3} \Upsilon^3 + \frac{1}{2} \kappa m \varphi \Upsilon \,, \quad
	\Upsilon \equiv | \varphi + \kappa | \,.
\label{superpotential_phi4_family_1}
\end{equation}
Here, $\kappa$ is an arbitrary real parameter.
One can equivalently rewrite it as
\begin{equation}
	\mathcal{W}_\kappa (\varphi) = \text{sign}( \varphi + \kappa ) \cdot \left( \frac{m^2}{ 4\lambda} \varphi  - \frac{\lambda}{3} \varphi^3 + \text{const} \right) \,.
\label{superpotential_phi4_family_2}
\end{equation}
Note that the symmetry under $\varphi \to -\varphi$ is lost for generic $\kappa$ (it is restored only for $\kappa$ either very large or vanishing).

It is straightforward to check that $ U = \frac{1}{2} | \mathcal{W}_\kappa ' |^2 $ identically for any $\kappa$.
Note that while the derivative $\mathcal{W}_\kappa '$ jumps at $\varphi = - \kappa$, its absolute value does not jump.
However, the second derivative $\mathcal{W}_\kappa ''$ really depends on $\kappa$ (note also that it will contain a $\delta$-function term).

Thus, we see that all theories with superpotentials \eqref{superpotential_phi4_family_1} share the same bosonic part \eqref{action_phi4_bos}, but differ in the fermion-boson interaction \eqref{action_phi4_ferm}.
When $|\kappa|$ is very large, such that we may take $\text{sign}( \varphi + \kappa )$ to be definite, the superpotential \eqref{superpotential_phi4_family_2} reduces to \eqref{superpotential_phi4_standard} up to an additive constant and an overall sign.
At intermediate $\kappa$, however, quantum dynamics depends on this new parameter.
In particular, for sufficiently small $\kappa$, the kink in this model is not BPS but rather semi-BPS; this was discussed in detail for the sphaleron in Sec.~\ref{sec:sphaleron_mstb}.

We leave more detailed investigation of this model to future work.

\section{More on central charges in two dimensions}
\label{sec:more_on_central_charges}

\subsection{ \boldmath ${\mathcal N}=(0,2)$ or $(2,0)$ }

This theory has complex supercharges $Q$ and $Q^\dagger$. They are chiral, which implies that at least some fields we deal with in such theories are massless.  
The ${\mathcal N}=(0,2)$ superalgebra has the form
\beq
\{ Q^\dagger Q\} = 2 (H\pm P)\,.
\eeq
{\em No central charges are possible}. ${\mathcal N}=(0,2)$ superspace as well as minimal and non-minimal Lagrangians are discussed, e.g., in \cite{Cui:2011uw}.

\subsection{ \boldmath ${\mathcal N}=(2,2)$ }

In this case we have four real supercharges, 
\beq
Q_\alpha^I\,\qquad \alpha =1,\,2 \,\, {\rm and}\,\, I=1,2\,.
\eeq
Their superalgebra takes the form
\begin{equation}
	\{Q_{\alpha}^{I},
	Q_{\beta}^{J}\}=2(\gamma^{\mu}\gamma^{0})_{\alpha\beta}P_{\mu}\delta^{IJ}+2i(\gamma^{5}\gamma^{0})_{\alpha\beta}Z^{\{IJ\}}+2i\gamma_{\alpha\beta}^{0}Z^{[IJ]} \,,
\label{45s}
\end{equation}
where $Z^{[IJ]}$ is antisymmetric in $I$, $J$ while $Z^{\{IJ\}}$ is 
symmetric.  The central charge $Z^{[IJ]}$ arises from the reduced momentum component $P_2$. This is the vortex central charge. The triplet $Z^{\{IJ\}}$
is the 3D projection of the 4D brane charge applicable to kinks (and domain lines in 3D).

Instead of (\ref{45s}) one can combine four real supercharges $Q_\alpha^I$ into two complex $\bar{Q}_\alpha$ and $Q_\alpha$, e.g.
\beq
Q_\alpha = Q^1_\alpha +i Q^2_\alpha\,,\quad \bar{Q}_\alpha =Q^1_\alpha -i Q^2_\alpha\,.
\eeq
In this notation the kink central charge appears in $\{Q_\alpha Q_\beta\}$ while the vortex CC in 3D shows up in $\{\bar Q_\alpha Q_\beta\}$.

\section{Note on a would-be enhanced supersymmetry}
\label{sec:almost_bps_2}

Here we would like to note an interesting property of the MSTB model from Sec.~\ref{sec:mstb} at a particular value $\kappa = 2$, also noted in \cite{AlonsoIzquierdo:2002pit}.
We stress that for any $\kappa > 1$ the model is still well-defined and has two degenerate vacua, but it does not have degenerate kinks --- there is only one kink interpolating between the two vacua.

Let us rewrite the bosonic potential \eqref{MSTB_potential} in a slightly different form:
\begin{equation}
	U(\phi_1,\phi_2) 
	= \frac{1}{2} \lambda^2 \abs{ (\phi_1 + i \phi_2)^2 - \frac{m^2}{ 4\lambda^2} }^2 + \left(  \frac{ \kappa^2  }{8} - \frac{1}{2} \right) m^2 \phi_2^2 \,.
\label{almost_bps_potential_1}
\end{equation}
When $\kappa = 2$, the last term vanishes.

Let us for a moment consider $\mathcal{N} = 2$ supersymmetric Wess-Zumino model with a single complex field $\Phi = \phi_1 + i \phi_2$ and the superpotential
\begin{equation}
	\mathcal{W}^{\mathbb{C}} = \frac{m^2}{ 4\lambda} \Phi - \frac{\lambda}{3} \Phi^3 \,.
\end{equation}
The scalar potential in this model is given by
\begin{equation}
	\abs{ \pdv{ \mathcal{W}^{\mathbb{C}} }{\Phi} }^2 
	= \lambda^2 \abs{ \Phi^2 - \frac{m^2}{ 4\lambda^2} }^2 \,.
\label{almost_bps_potential_2}
\end{equation}
One can see that the bosonic potentials in Eqs.~\eqref{almost_bps_potential_1} and \eqref{almost_bps_potential_2} are proportional to each other at $\kappa = 2$. 

Thus, at $\kappa = 2$, the bosonic part of the MSTB model coincides with the bosonic part of an $\mathcal{N}=2$ theory.
The fermionic part is, of course, different, as the superpotential for the MSTB model \eqref{MSTB_superpotential} is still non-holomorphic for any value of $\kappa$.

%\section{Large figures}
%\label{sec:lg_fig}
%
%
%
%
%\clearpage

\bibliographystyle{JHEP}

\bibliography{main}

\end{document}